\documentclass[12pt]{article}
\pdfoutput=1
\usepackage{latexsym, graphicx} 
\usepackage{amsmath}
\usepackage{amssymb}
\usepackage{caption}
\usepackage{subcaption}
\usepackage{tensor}
\usepackage{cite}
\usepackage[utf8]{inputenc}
\numberwithin{equation}{section}

\usepackage[colorlinks=true, linkcolor=blue, bookmarks=true]{hyperref}
\newcommand{\arXiv}[1]{\href{http://www.arXiv.org/abs/#1}{#1}}

\renewcommand{\title}[1]{\vbox{\center\bf{\Large{#1}}}\vspace{5mm}}
  \renewcommand{\author}[1]{\vbox{\center#1}\vspace{5mm}}
  \newcommand{\address}[1]{\vbox{\center\em#1}}

\begin{document}
\begin{titlepage}
\begin{center}
\hfill \\
\hfill \\
\vskip 1cm
\title{Superluminal chaos after a quantum quench}
\author {Vijay~Balasubramanian,$^{1,2}$ Ben~Craps,$^{2}$ Marine~De Clerck,$^{2}$ K\'evin~Nguyen$^{2}$\vspace{2mm}}
\address{
$^{1}$David Rittenhouse Laboratory, University of Pennsylvania, \\ Philadelphia, PA 19104, USA \\
$^{2}$Theoretische Natuurkunde, Vrije Universiteit Brussel (VUB) and\\
The International Solvay Institutes, Pleinlaan 2, B-1050 Brussels, Belgium \vspace{1mm}\\}
\end{center}

\begin{abstract}
Thermal states holographically dual to black holes in Einstein gravity display maximal Lyapunov growth as well as ``butterfly effect cones''. We study these effects in highly non-equilibrium states, obtained from an initial thermal state by the sudden injection of energy. We do this by computing out-of-time-order correlators (OTOCs) in BTZ-Vaidya spacetimes, which describe transitions between black holes at different temperatures. If both pairs of boundary operators appearing in the OTOC are inserted before the energy injection, we recover standard results, with butterfly effect cones displaying a light-cone structure. But when one pair of operators is inserted before and the other pair after the energy injection, the Lyapunov growth saturates the chaos bounds set by the local temperatures and the butterfly effect cones can ``open up", becoming superluminal, albeit in a way that does not violate causality. In the limiting  case, in which the initial state is the vacuum, Lyapunov growth only starts after the energy injection. Our computations of the OTOCs are phrased in terms of gravitationally interacting particles, where fields are treated in a geodesic approximation and the eikonal phase shift is expressed in terms of stress tensors and shock waves associated to geodesics. 

\end{abstract}
\end{titlepage}	

\section{Introduction}
In recent years, a connection between chaotic behavior of strongly coupled holographic theories and gravitational physics in AdS has been identified, leading to the conjecture that thermal states with holographic duals are the most chaotic systems \cite{Sekino:2008he}. This idea was made precise in the context of a quantum mechanical bound on chaos \cite{Maldacena:2015waa}, which is saturated by thermal CFT states dual to AdS black holes \cite{Shenker:2013pqa,Shenker:2013yza,Roberts:2014isa,Shenker:2014cwa,Roberts:2014ifa}. 

In thermal systems at inverse temperature $\beta$, chaos can be diagnosed by a specific type of out-of-time-order correlator (OTOC)
\begin{equation}
\langle V(0)W(t)V(0)W(t)\rangle_\beta
\label{otoc1}
\end{equation} 
between ``simple'' hermitian operators $V$ and $W$ (which can be localized in spatially extended systems, or involve a small number of degrees of freedom in matrix models) \cite{Kitaev}.
In a chaotic system, the OTOC is expected to vanish at late times, independently of the choice of operators $V$ and $W$ \cite{Shenker:2014cwa}. 

The OTOC is closely related to the  commutator squared (CS), 
\begin{equation}
\label{commutator_squared}
C(t)\equiv -\langle \left[W(t),V(0)\right]^2\rangle_{\beta},
\end{equation}
which is used as a diagnostic for the presence of chaos in quantum systems because of its relation to the classical Poisson bracket and the exponentially divergent trajectories expected in classical chaos \cite{Larkin,Kitaev}.  A quantum system is said to be chaotic if the commutator squared grows exponentially in time 
\begin{equation}
-\langle \left[W(t),V(0)\right]^2\rangle \sim \frac{1}{N^2}\ e^{2\lambda_L t},
\end{equation}
with characteristic Lyapunov exponent $\lambda_L$ and where $N$ corresponds to the number of degrees of freedom of the system (we restrict our attention to systems without spatial extent for the moment). More precisely,  this exponential behavior, which is related to the phenomenon of \textit{fast scrambling}, is expected for times $t_d \ll t \ll t_*$, where $t_d \sim \beta$ is the dissipation time and $t_* \sim \lambda_L^{-1} \log N$ the scrambling time \cite{Sekino:2008he}. The scrambling time gives the time scale needed for a perturbation involving a few degrees of freedom to spread among all the degrees of freedom of the system \cite{Sekino:2008he,Shenker:2013pqa}. Any operations performed after the scrambling time on a subset of the degrees of freedom will be unable to reconstruct the information of the perturbation. 

The connection between the OTOC (\ref{otoc1}) and the squared commutator  becomes clear after expanding \eqref{commutator_squared}
\begin{align}
-\langle \left[W(t),V(0)\right]^2 \rangle = \langle WVVW \rangle + \langle VWWV \rangle-2\ \text{Re} \langle VWVW \rangle,
\end{align}
and noting that the first two terms factorize in a thermal state after a dissipation time.  For example, 
\begin{equation}
\langle V(0) W(t) W(t) V(0) \rangle_\beta \approx \langle W(t) W(t) \rangle_\beta\ \langle V(0) V(0) \rangle_\beta+\mathcal{O}(e^{-t/t_d}),
\end{equation}
since it can be thought of as the two-point function of $W(t)$ in the thermal state perturbed by the insertion of operator $V(0)$ \cite{Roberts:2014ifa}. If the energy injected by V is small, the state will relax and the expectation value will approach the thermal value multiplied by the norm of the state \cite{Roberts:2014ifa}, essentially because all time-ordered correlations decay to products of expectation values in a thermal state after the dissipation time \cite{Ruelle}.   This information can be used to simplify \eqref{commutator_squared} in the regime $t\gg t_d$,
\begin{equation}
D(t)\equiv \frac{-\langle \left[W(t),V(0)\right]^2 \rangle }{\langle WW\rangle \langle VV \rangle}\approx 2-2\ \frac{\text{Re} \langle V(0)W(t)V(0)W(t) \rangle}{\langle WW\rangle \langle VV \rangle}.
\label{normalized CS}
\end{equation}
Thus, the normalized commutator squared $D(t)$ is simply related to the OTOC after the dissipation time in thermal systems. In addition, the characteristic late-time vanishing of the OTOC for chaotic systems is equivalent to saturation of the normalized commutator squared after the exponential Lyapunov regime.  

The above considerations apply for quantum systems without a notion of spatial locality. If Hamiltonian interactions are local in space, i.e., if they decay for operators that are spatially separated, then there are two scrambling time scales. The first is associated to \textit{fast} scrambling among local degrees of freedom and can be characterized as described above. The second type of scrambling is associated to spreading of quantum information in space and in some cases, a \textit{butterfly velocity} $v_B$ can be associated to this propagation \cite{Roberts:2014isa}, roughly speaking measuring the rate of growth of the region that must be tracked to reconstruct an original perturbation with a given accuracy.  In nonrelativistic systems the butterfly velocity is bounded by Lieb-Robinson bounds constraining the growth of commutators in discrete Hamiltonian systems to be at most ballistic \cite{Lieb:1972wy,Hastings:2005pr}.    In relativistic theories, it would be natural to expect that the butterfly velocity would be bounded by the speed of light. Note though that if the region that must be measured to reconstruct an original perturbation grows faster than light, it does not imply that this can be used to send information faster than light. Consider an astronaut on the moon who keeps track of laser dots of various colors originating from Earth; these dots can move faster than light on the surface of the moon, and the area needed to keep track of all of them can also grow faster than light. But of course this does not allow astronauts on the moon to transmit information faster than light.

Unlike in classical systems, there is a bound on the Lyapunov exponent in quantum mechanics. For thermal systems, the bound is simply set by the temperature \cite{Maldacena:2015waa},
\begin{equation}
\label{Lyapunov_bound}
\lambda_L \leq \frac{2\pi}{\beta},
\end{equation}
while some generalizations thereof have been recently studied in \cite{Mezei:2019dfv,Halder:2019ric}. This bound is actually saturated in conformal field theories (CFTs) by thermal states dual to AdS black holes. The OTOC of such thermal CFT states at inverse temperature $\beta$ can be computed in two distinct ways. In \cite{Roberts:2014ifa}, the correlator was computed using CFT methods, by taking advantage of the exponential map between a thermal state and the vacuum state and making use of vacuum block techniques. In the gravitational bulk interpretation \cite{Shenker:2014cwa}, the chaotic behavior can be understood in terms of a more conventional time-ordered high-energy bulk scattering of particle excitations created by the insertion of the operators $V$ and $W$. In this picture, the OTOC is the overlap of bulk \textit{in}- and \textit{out}-states created by the boundary insertions. This can be seen by first viewing the OTOC as the overlap $\langle \Psi'|\Psi\rangle$ of the two states
\begin{equation}
|\Psi\rangle\equiv V(t_-,0) W(t_+,x)|0\rangle \qquad \text{and} \qquad |\Psi'\rangle \equiv W(t_+,x)V(t_-,0)|0\rangle,
\end{equation}
with $|0\rangle$ the state dual to the background bulk geometry with no excitations of the $W, V$ operators, and $t_- < t_+$.
\begin{figure}[h!]
	\centering
	\begin{subfigure}[b]{0.49\textwidth}
		\centering
		\includegraphics[scale=0.6]{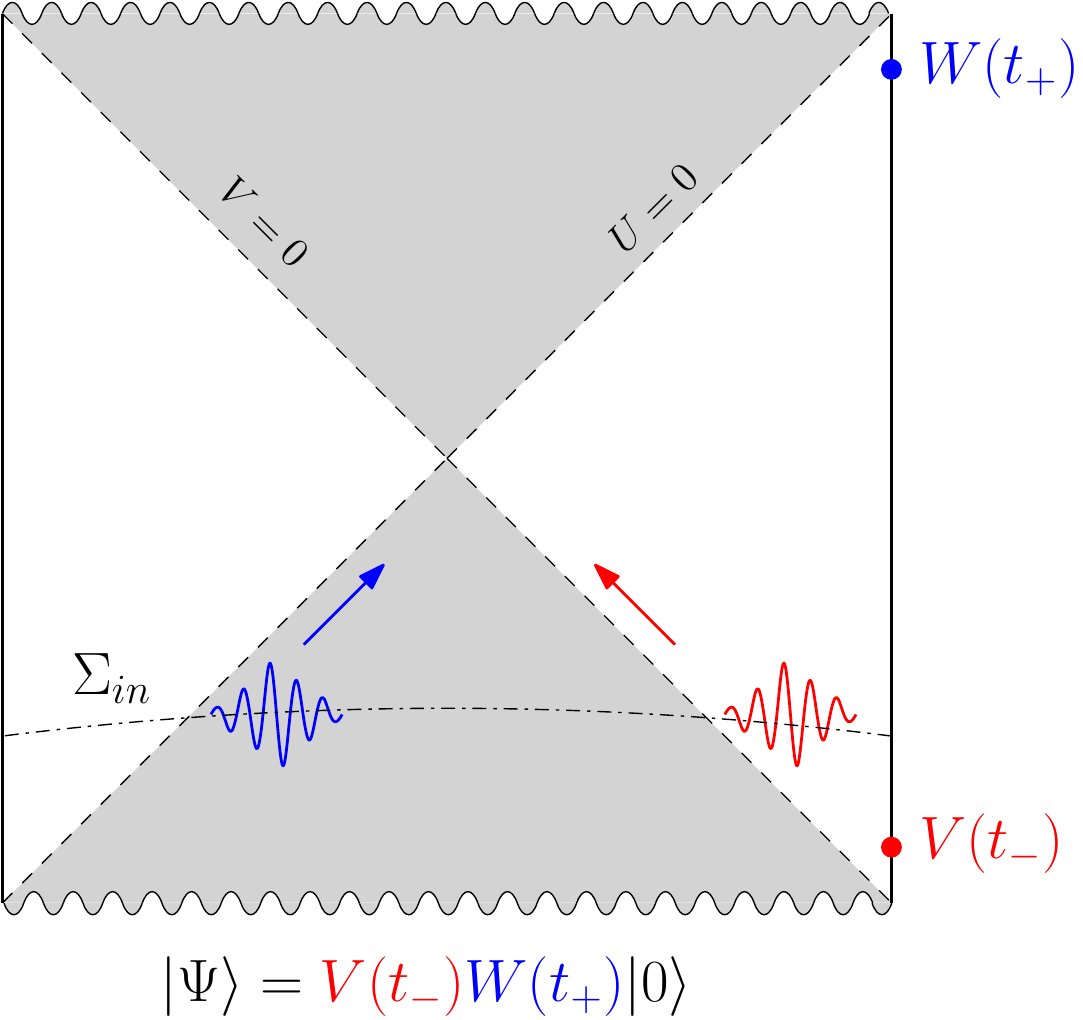}
		\vskip 2mm
		\caption{}
		\label{fig: intro btza}
	\end{subfigure}
	\begin{subfigure}[b]{0.49\textwidth}
		\centering
		\includegraphics[scale=0.6]{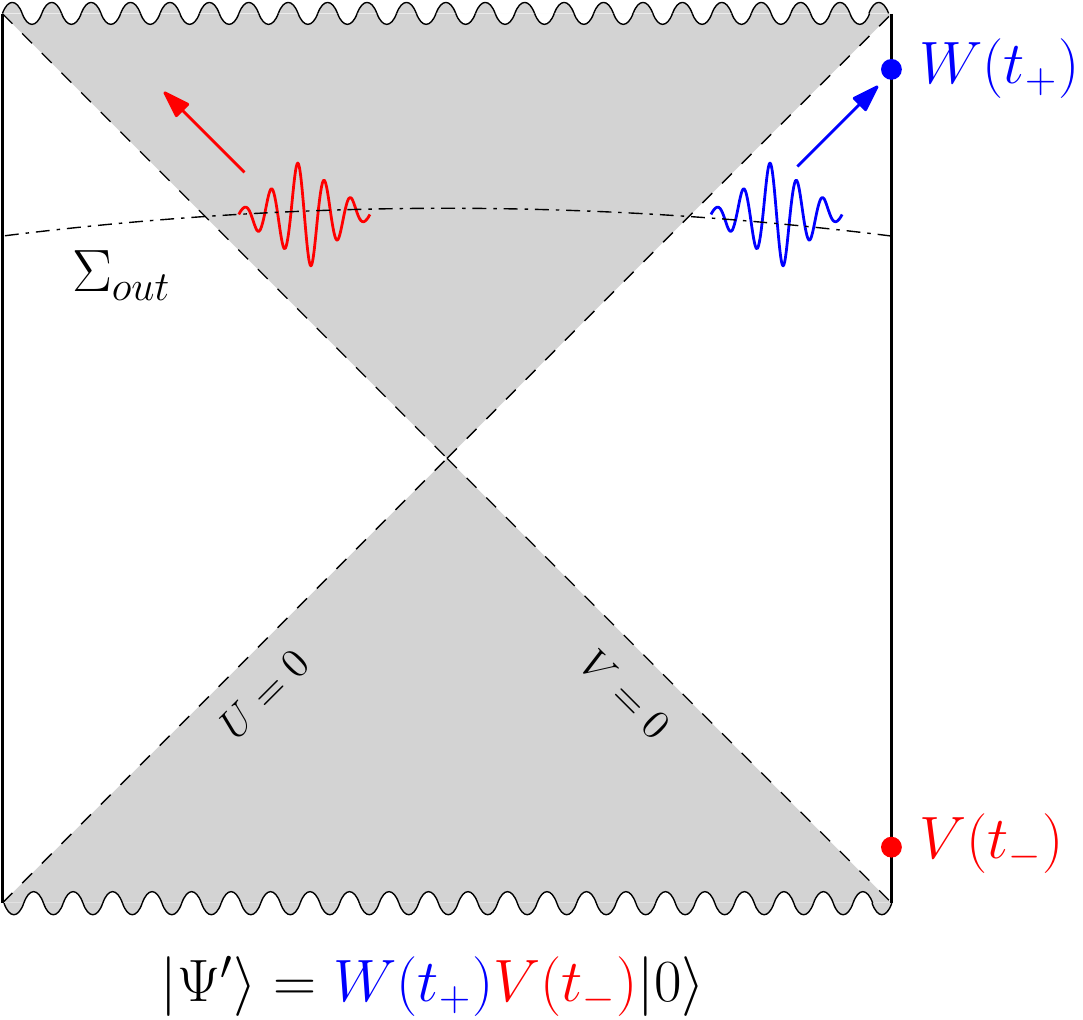}
		\caption{}
		\vskip 2mm
		\label{fig: intro btzb}
	\end{subfigure}
	\begin{subfigure}[b]{0.49\textwidth}
		\centering
		\includegraphics[scale=0.6]{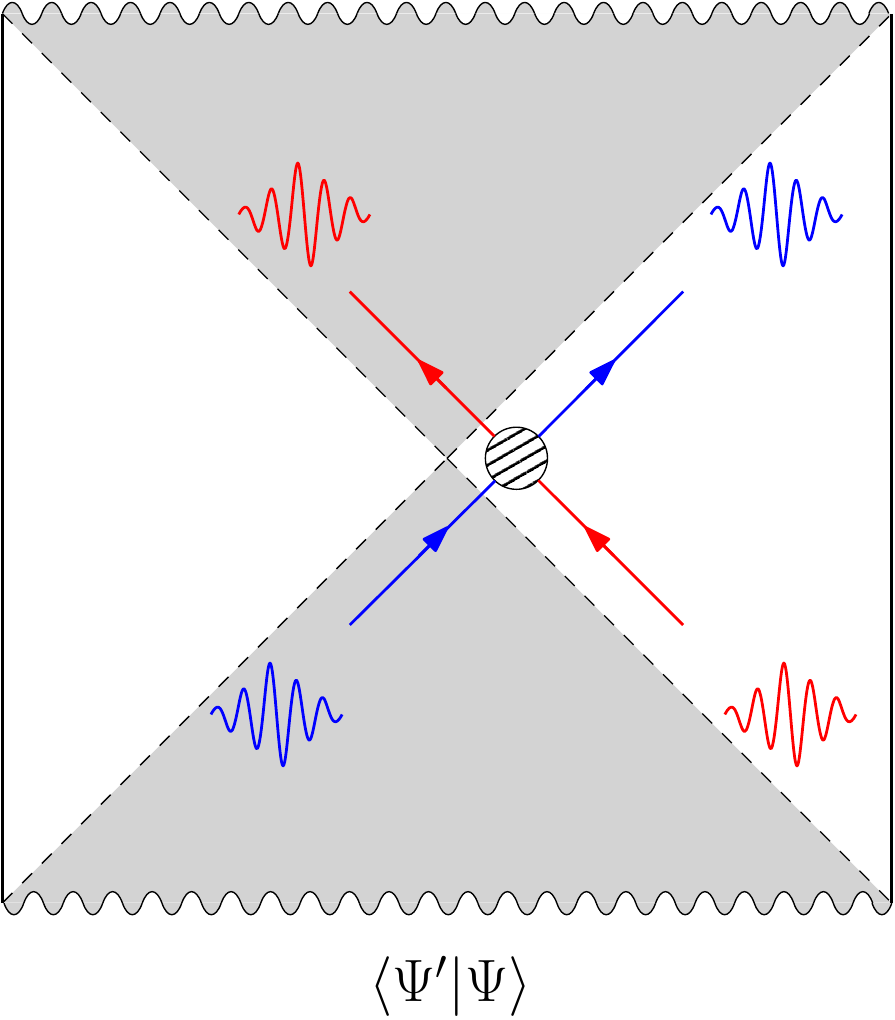}
		\vskip 2mm
		\caption{}
		\label{fig: intro btzc}
	\end{subfigure}
	\caption{(a) The bulk picture of the in-state. This state has a simple interpretation at early times. (b) The bulk picture of the out-state. This state has a simple interpretation at late times. (c) The bulk picture of the OTOC is the overlap of the in-state (a) and the out-state (b), which corresponds to bulk scattering. Figure adapted from \cite{Shenker:2014cwa}.}
	\label{fig: intro btz}
\end{figure}
The state $|\Psi\rangle$ can be understood as follows. As a first step, $W$ is inserted at the boundary of the empty geometry at time $t_+$. In order to apply the operator $V$ at time $t_-$ on this state, the perturbation $W$ is evolved backwards in time using the bulk-to-boundary propagator. Subsequently, the $V$ perturbation is inserted.  The state as a whole has a simple bulk picture at early times (Fig.~\ref{fig: intro btza}), and is therefore interpreted as an in-state. Expressing the same state at late times would require solving an involved scattering problem. A similar picture can be obtained for $ |\Psi'\rangle$, where the operators are inserted in the opposite order. The $V$ operator is inserted at time $t_-$ on the empty background geometry and is evolved forward in time in order to apply the $W$ operator at a later time $t_+$. The $V$ perturbation creates field excitations that fall into the black hole, and the $W$ operator can be applied without disturbing the $V$ field excitations too much. This state has a simple form at late times (Fig.~\ref{fig: intro btzb}), and is interpreted as an out-state. A Fourier decomposition into momentum states allowed to derive a formula for the OTOC that describes the correlator as a $2\to 2$ bulk scattering of highly energetic particles created by the perturbations $V$ and $W$ \cite{Shenker:2014cwa}. This scattering is illustrated in Fig.~\ref{fig: intro btzc}.

In both methods, the OTOC yields
\begin{equation}
\label{OTOC_shenker}
\frac{\langle V(t_-,0) W(t_+,x) V(t_-,0) W(t_+,x) \rangle}{\langle VV \rangle\ \langle WW \rangle}= \left(1-\frac{8\pi i G_N \Delta_W}{\varepsilon^2} e^{\frac{2\pi}{\beta}(t_+-t_--|x|)}\right)^{-\Delta_V},
\end{equation}
under the assumption $\Delta_W \gg \Delta_V$ and where the parameter $\varepsilon$ has been introduced to regulate the divergences associated to the coincident insertions of operators. For a small enough separation $t_+-t_--|x|$ between pairs of operator insertions, the commutator squared \eqref{normalized CS} displays an exponential regime,
\begin{equation}
\label{Lyapunov}
D(t_+-t_-,x) \approx \left(\frac{8\pi G_N \Delta_W\Delta_V}{\varepsilon^2} e^{\frac{2\pi}{\beta}(t_+-t_--|x|)}\right)^2,
\end{equation}
with Lyapunov exponent $\lambda_L=\frac{2\pi}{\beta}$. It saturates the bound \eqref{Lyapunov_bound} and hence supports the idea that black holes are the fastest possible scramblers. 

In the holographic bulk interpretation, the Lyapunov behavior arises from the blueshift that such particles are subjected to, due to the gravitational acceleration in the black hole background. Initially small perturbations around the black hole background end up energetic enough to create strong gravitational fields known as \textit{shock waves}, causing them to interact strongly with other particles travelling near the black hole. This phenomenon is at the core of the chaotic behavior of the dual holographic CFT.
From \eqref{Lyapunov}, the spatial region for which $D(t_+-t_-,x)$ is non-negligible can be seen to grow ballistically with butterfly velocity $v_B = 1$,
\begin{equation}
D(t_+-t_-,x) \sim 1 \qquad \Leftrightarrow \qquad x \sim t_+-t_- \equiv  v_B\ \left(t_+-t_- \right).
\end{equation}

In addition, expression \eqref{OTOC_shenker} also encodes the decay of the OTOC at very late times $t\gg \frac{\beta}{2\pi} \ln G_N^{-1}$, 
\begin{equation}
\frac{\langle V(t_-,0) W(t_+,x) V(t_-,0) W(t_+,x) \rangle}{\langle VV \rangle\ \langle WW \rangle}\approx  e^{-\frac{2\pi}{\beta}\Delta_V(t_+-t_--|x|)}.
\end{equation}
This decay is controlled by the lowest quasinormal mode frequency $\omega_{QN}=\frac{2\pi}{\beta} \Delta_V$ associated to the light operator $V$ of conformal dimension $\Delta_V$ \cite{Birmingham:2001pj}. This shows that dissipative effects eventually take over the chaotic Lyapunov growth.
\begin{figure}[h!]
	\centering
	\includegraphics[height=8cm]{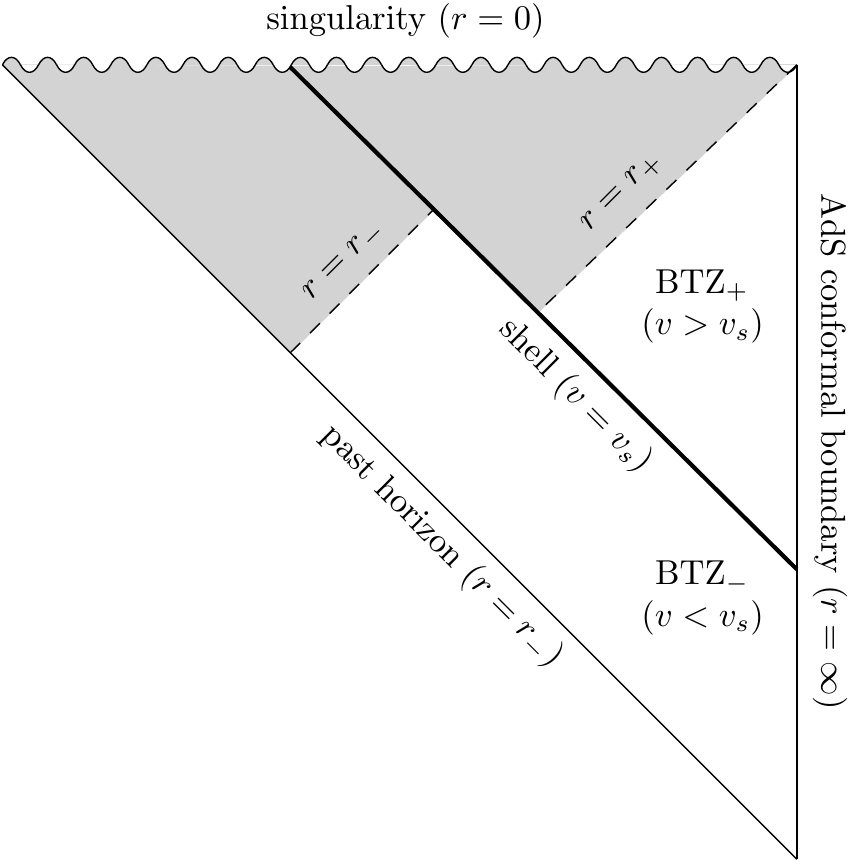}
	\caption{Penrose diagram of BTZ-Vaidya.}
	\label{fig:Penrose}
\end{figure}

In this paper, we investigate the chaotic behavior of a system in the process of thermalization by holographically computing OTOCs in a thermal state undergoing a global quench. On the gravity side, this state corresponds to an initial BTZ black hole with inverse temperature $\beta_-$ (BTZ$_-$) which, after a global energy injection at the boundary of the spacetime that creates an infalling shell, turns into a black hole with inverse temperature $\beta_+ < \beta_-$ (BTZ$_+$), as depicted in Fig.\ \ref{fig:Penrose}. These bulk states are known as BTZ-Vaidya spacetimes. 

We first extend the computation of the OTOC in the eternal BTZ black hole, which was reviewed above, to the case of dynamically growing (forming) black holes. As for the eternal black hole, boundary perturbations can be propagated into the bulk and the OTOC can be interpreted as an overlap of an in- and an out-state.
\begin{figure}[h!]
	\centering
	\begin{subfigure}[b]{0.49\textwidth}
		\centering
		\includegraphics[scale=0.8]{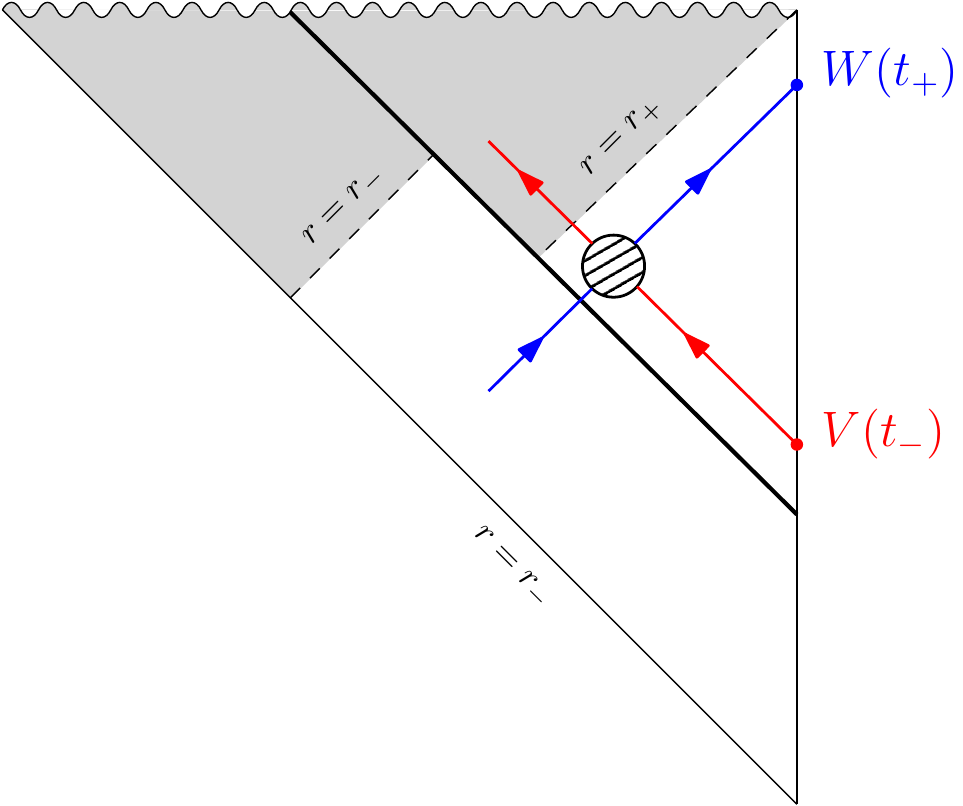}
		\vskip 2mm
		\caption{}
		\label{fig: intro btz_vaidyaa}
	\end{subfigure}
	\begin{subfigure}[b]{0.49\textwidth}
		\centering
		\includegraphics[scale=0.8]{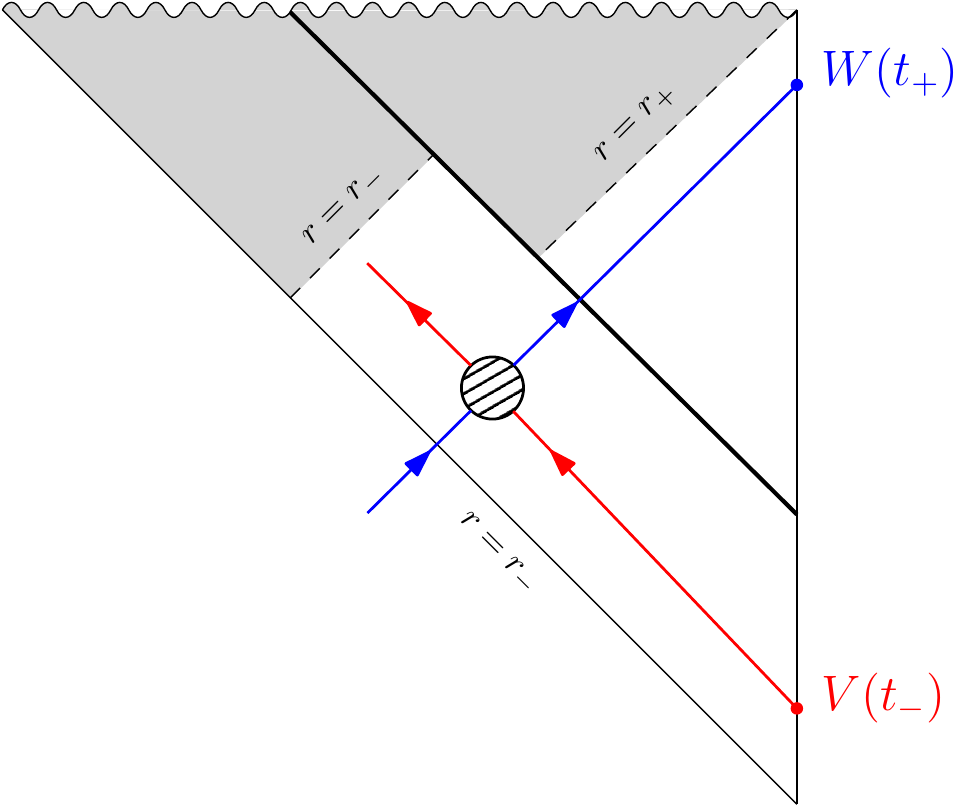}
		\caption{}
		\vskip 2mm
		\label{fig: intro btz_vaidyab}
	\end{subfigure}
	\caption{(a) When all the operators are inserted after the quench, the situation is very similar to the eternal BTZ case, as the scattering happens in the BTZ$_+$ part of the spacetime. (b) When one pair of operators is inserted before and the other after the shell, the scattering typically happens in BTZ$_-$, which modifies the behavior of the OTOC.}
	\label{fig: intro btz_vaidya}
\end{figure}
When all operators are inserted before (after) the shell, the shock-wave scattering process happens in the BTZ$_-$ (BTZ$_+$) part of the spacetime, as can be seen in Fig.~\ref{fig: intro btz_vaidyaa}. In this case, we recover the standard Lyapunov behaviors in thermal states, but we are able to extend previous results to time separations between the inserted operators of the order of the dissipation time and smaller. 

However, when one pair of operators is inserted before and the other after the quench, the shock-wave scattering process often takes place in BTZ$_-$, as illustrated on Fig.~\ref{fig: intro btz_vaidyab} (this happens if the transverse separation between the operators is sufficiently small; the precise condition will be given in Section \ref{section:eikonal}).\footnote{Note that the scattering process is not just between the inserted matter particles, but between these particles and the extended gravitational shock waves that they create -- thus the dominant scattering can occur in different parts of the geometry depending on the details.}  This fact modifies the (spatial) behavior of the OTOC, as compared to the thermal equilibrium case.  There is an unexpected consequence: right after the quench, the butterfly velocity is set by the ratio of the second to the first horizon radius, and is superluminal. However, as we will see, this effect relaxes before it can violate causality and the overall spread remains within the lightcone. Moreover, two exponential behaviors develop with separate Lyapunov exponents, set by each of the temperatures, that multiply the time separation between the insertion of the quench and the insertion of each of the operators in the part of the spacetime associated to that temperature. From the holographic perspective, this feature is expected in view of the blueshift that particles experience before the scattering, which is the origin of the Lyapunov exponents in black hole geometries. By taking the zero-temperature limit of the initial state ($\beta_- \rightarrow \infty$), we extend our results to CFT states which start in the vacuum and thermalize after a global quench, allowing us to study the onset of chaos in dynamically forming black holes.   As a cross-check we also recover the vacuum CFT result obtained by \cite{Roberts:2014ifa}, for all the operators inserted before the quench. 

The plan of the paper is as follows. In Section~\ref{section:OTOC formula}, we derive for a general spacetime a position-space WKB analogue of the overlap integral that computes the OTOC  in \cite{Shenker:2014cwa}.
We apply this formula to BTZ-Vaidya spacetimes in Section~\ref{section:geodesics} and discuss our results in Section~\ref{section:results}. The appendices contain technical material regarding the properties of geodesics, the WKB approximation and shock waves.

\section{OTOC in asymptotically AdS spacetimes}\label{section:OTOC formula}
Let us consider an asymptotically AdS background spacetime on which two massive Klein-Gordon real scalar fields $\phi_V$ and $\phi_W$ propagate. We will derive a general formula for the out-of-time-order correlator (OTOC)
\begin{equation}
\label{OTOC}
\langle \phi_V(X_1) \phi_W(X_2) \phi_V(X_3) \phi_W(X_4) \rangle,
\end{equation}
where the insertion points $X_2, X_4$ lie in the future of $X_1, X_3$ or are spacelike-separated from them.  As the radial location of the  insertion points $X_i$ are taken to the boundary of AdS space, (\ref{OTOC}) computes (after a radial rescaling) the associated correlator in the dual conformal field theory defined on the boundary \cite{Banks:1998dd,Balasubramanian:1998de,Balasubramanian:1998sn,Balasubramanian:1999ri}.   We will carry out all our calculations in such a near-boundary limit of the insertion points so that the dependence of the correlation functions on time and the transverse space can be interpreted as also computing these dependences in the dual field theory.

 In the limit that all insertion points are close to the conformal boundary, one can make use of the WKB (geodesic) approximation for propagators of the scalar fields. Indeed, in any asymptotically AdS space, a timelike geodesic gets more energetic as one of its endpoints gets closer to the conformal boundary. This is exactly the regime in which we expect the WKB approximation to be reliable.\footnote{See \cite{Balasubramanian:1999zv} for related remarks in the context of propagators for heavy fields in AdS space.}   In addition, at leading order in an inverse energy expansion $e^{-1}\ll 1$, such highly energetic geodesics follow light rays and their gravitational backreaction can be exactly computed in the form of \textit{shock waves}. It will hence be useful to equip the considered background spacetime with coordinates $X=(u,v,x)$, where $u,v$ are advanced and retarded lightlike coordinates and $x$ labels the transverse direction.\\

\begin{figure}[h!]
	\centering
	\begin{subfigure}[b]{0.49\textwidth}
	\includegraphics[scale=0.85]{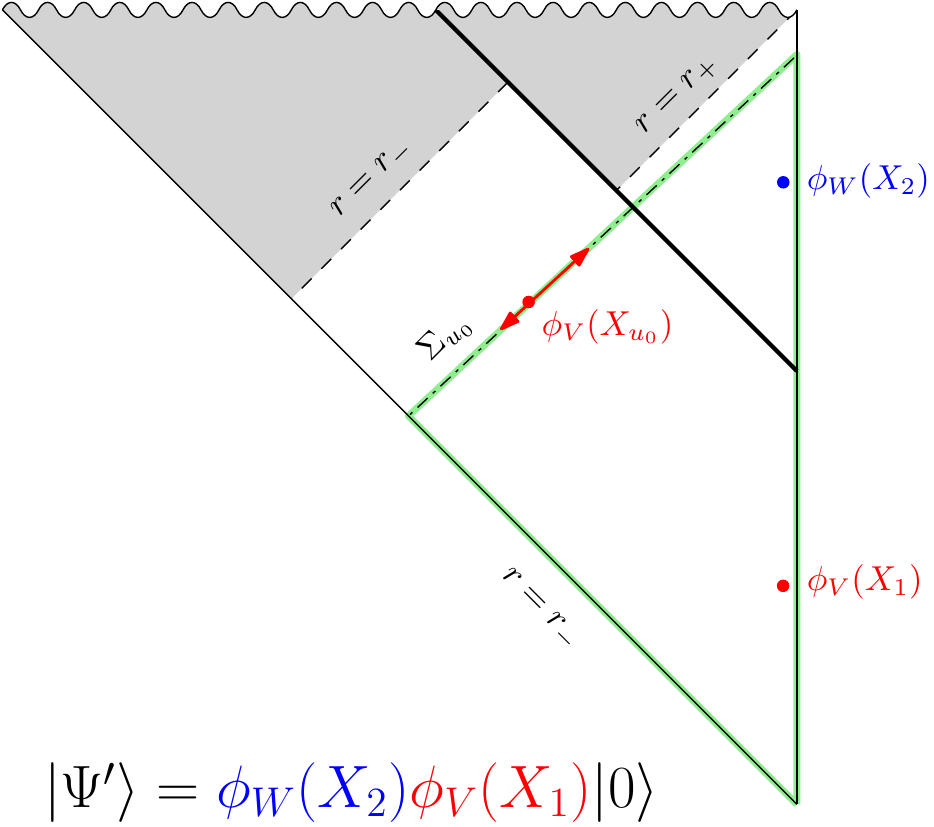}
	\caption{}
	\label{fig:Psi' Vaidya}
	\end{subfigure}
	\begin{subfigure}[b]{0.49\textwidth}
	\includegraphics[scale=0.85]{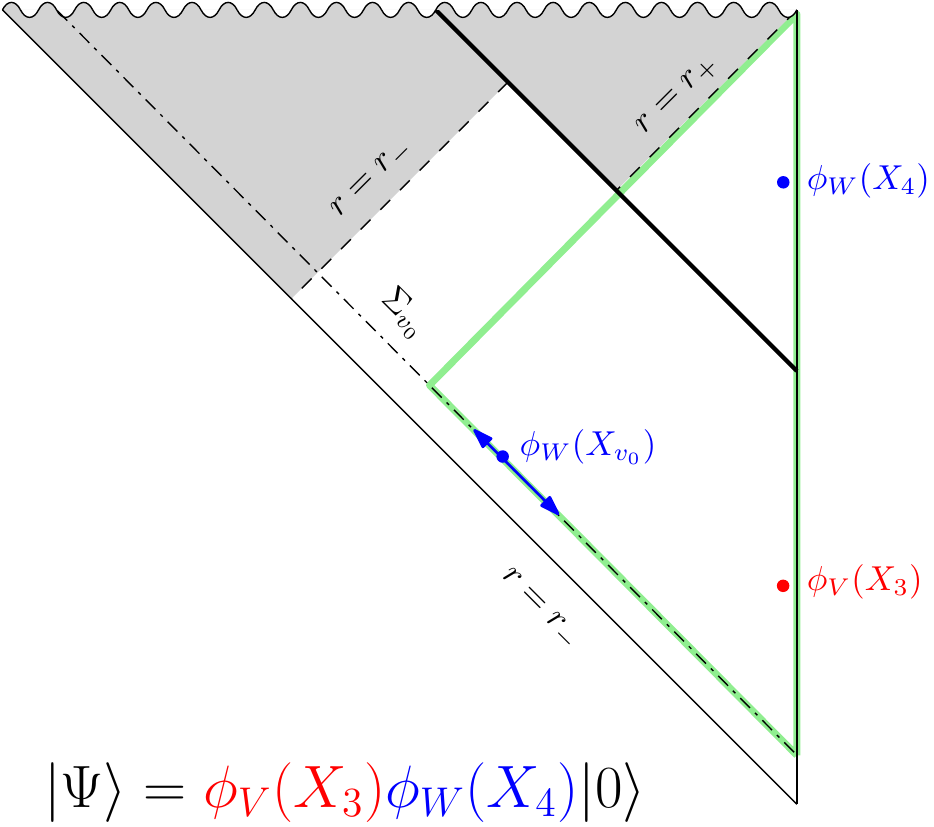}
	\caption{}
	\label{fig:Psi Vaidya}
	\end{subfigure}
	\vskip 5mm
	\begin{subfigure}[b]{0.49\textwidth}
	\includegraphics[scale=0.85]{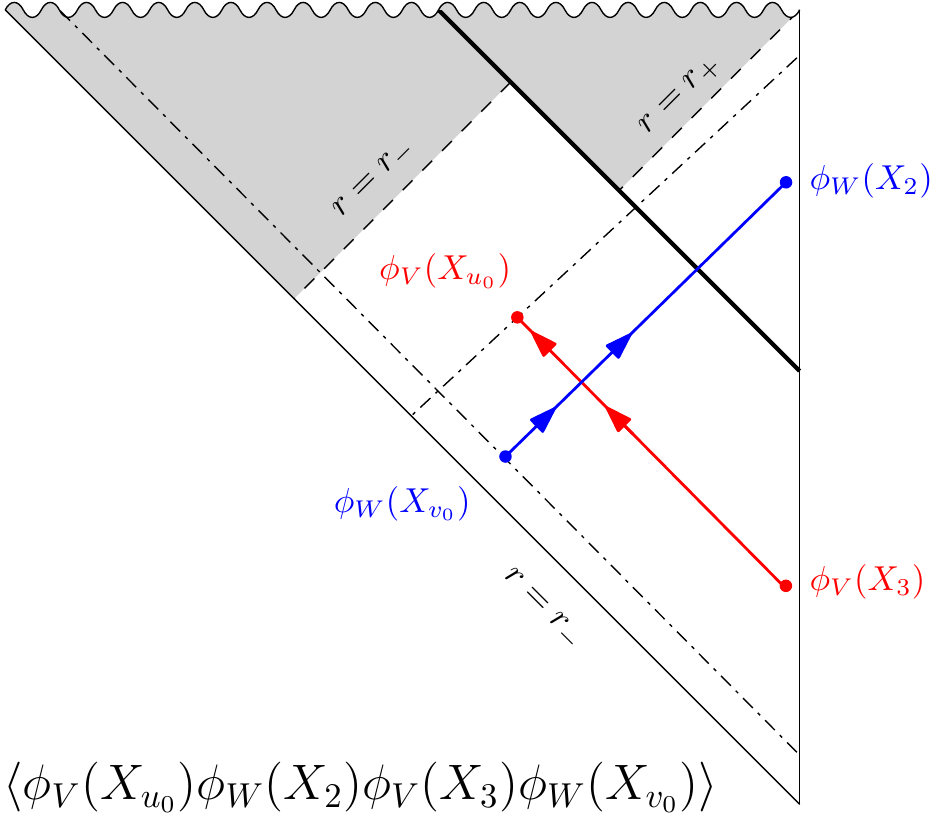}
	\caption{}
	\label{}
	\end{subfigure}
	\caption{(a) The operator $\phi_V(X_1)$ is represented on a future null slice $\Sigma_{u_0}$ in terms of the operator $\phi_V(X_{u_0})$ integrated over the point $X_{u_0} \in \Sigma_{u_0}$. The enclosing surface $\partial \mathcal{V}$ is indicated in light green. (b) The operator $\phi_W(X_4)$ is similarly represented on a past null slice $\Sigma_{v_0}$ in terms of the operator $\phi_W(X_{v_0})$ integrated over the point $X_{v_0} \in \Sigma_{v_0}$. (c)~The overlap $\langle \Psi'|\Psi\rangle$ only involves time-ordered four-point correlators such as $\langle \phi_V(X_{u_0}) \phi_W(X_2) \phi_V(X_3) \phi_W(X_{v_0})\rangle$. Their evaluation boils down to the computation of highly energetic geodesics indicated by the straight colored arrows, and their associated gravitational fields. As shown in the next sections, we anticipate that the overlap $\langle \Psi'|\Psi\rangle$ is dominated by the contribution of a single pair of such geodesics.}
	\label{fig:surfaces}
\end{figure}

We want to find a formula applying to arbitrary asymptotically AdS spacetimes. In the general case, it is not clear whether the method of \cite{Shenker:2014cwa} may be directly applied, especially in time-dependent backgrounds for which asymptotic \textit{in} and \textit{out} Hilbert spaces are not always available. However, the intuition that the OTOC is dominated by the scattering of highly energetic particles \cite{videomaldacena} should still hold.\footnote{This argument relies on the validity of the geodesic approximation to field propagators, and hence on the existence of highly energetic geodesics. This may happen either due to gravitational attraction by a black hole, or simply due to the gravitational potential barrier near the boundary of any asymptotically AdS spacetime.} In this section, treating field excitations in the WKB (geodesic) approximation, we obtain a general position-space formula for the OTOC \eqref{OTOC} in terms of the scattering of highly energetic geodesics or point-like particles. We will proceed as follows, viewing the OTOC as the overlap $\langle \Psi'|\Psi\rangle$ of the two states
\begin{equation}
|\Psi\rangle\equiv \phi_V(X_3) \phi_W(X_4)|0\rangle \qquad \text{and} \qquad |\Psi'\rangle \equiv \phi_W(X_2)\phi_V(X_1)|0\rangle.
\end{equation}
First, we give a simple representation of $|\Psi \rangle$ at earlier times and of $|\Psi' \rangle$ at later times, in the spirit presented in the Introduction. For this, we use Green's second identity to represent $\phi_V(X_1)$ on a future null slice $\Sigma_{u_0}$ and $\phi_W(X_4)$ on a past null slice $\Sigma_{v_0}$. As a result, the overlap is purely expressed in terms of time-ordered four-point correlators (Fig.~\ref{fig:surfaces}). Second, we write down the gravitational path integral representation of these time-ordered four-point correlators. Treating field propagators in the WKB approximation and taking the classical gravity limit, we show that their evaluation boils down to the computation of classical quantities: highly energetic geodesics and their associated \textit{shock wave} gravitational fields. The method developed here shares many similarities with \cite{tHooft:1990fkf,Kiem:1995iy,Kabat:1992tb,Balasubramanian:1995sm}. 

\subsubsection*{From out-of-time-order to time-ordered correlators}
We first consider the state $|\Psi'\rangle$. We want to express the field operator $\phi_V$ inserted at $X_1$ in terms of the same operator inserted at points $X_{u_0}=(u_0,v,x)$ belonging to a null slice $\Sigma_{u_0}$ of constant $u=u_0$ such that any $X_{u_0} \in \Sigma_{u_0}$ is either in the future of $X_2, X_4$ or spacelike-separated from them. For this we first choose a volume $\mathcal{V}$ containing all insertions, which is enclosed by the union of $\Sigma_{u_0}$, parts of the conformal boundary and of the past horizon, as well as transverse surfaces of constant $x= x_{\pm \infty}$ being infinitely far away from all operator insertions. The projection of $\partial \mathcal{V}$ onto the $(u,v)$-plane is indicated in light green in Fig.~\ref{fig:Psi' Vaidya}.  We then apply Green's second identity,
\begin{equation}
\int_{\mathcal{V}} d^{d+1}X\ \sqrt{-g} \left(\psi_1 \square \psi_2- \psi_2 \square \psi_1\right)= \oint_{\partial \mathcal{V}} d\Sigma^\mu \left(\psi_1 \partial_\mu \psi_2- \psi_2 \partial_\mu \psi_1\right), 
\end{equation}
to the following solutions of the Klein-Gordon equation,
\begin{align}
\psi_1(X)&=\phi_V(X),\\
\psi_2(X)&=G^R_V\left(X,X_1\right).
\end{align}
The retarded Green function $G^R_V(X,Y)$ vanishes for $(X-Y)^2>0$ and for $X^0<Y^0$, and satisfies
\begin{equation}
\left(\square -m_V^2\right) G^R_V(X,Y)=\frac{\delta^{(d+1)}(X-Y)}{\sqrt{-g}}.
\end{equation}
We consider Dirichlet boundary conditions at the conformal boundary\footnote{In the standard AdS/CFT terminology, non-normalizable sources for the bulk fields $\phi_V$ and $\phi_W$ are turned off.} and no incoming initial conditions at the past horizon for the scalar field $\phi_V$ such that $G^R_V\left(X;X_1\right)$ vanishes there.  On the enclosing surface $\partial \mathcal{V}$, this Green function has therefore support only on $\Sigma_{u_0} \subset \partial \mathcal{V}$ and we get 
\begin{align}
\label{hat phi}
\hat{\phi}_V(X_1)&=\int_{u=u_0} d\Sigma^v \left(\hat{\phi}_V(X_{u_0}) \stackrel{\leftrightarrow}{\partial_v} G^R_V\left(X_{u_0},X_1\right)\right),
\end{align} 
giving a representation of $\hat{\phi}_V(X_1)$ in terms of fields on the null surface  $\Sigma_{u_0}$.   We can now write the state $|\Psi'\rangle$ by simply multiplying the right hand side of (\ref{hat phi}) by $\hat{\phi}_W(X_2)$: 
\begin{align}
\label{Psi'}
|\Psi'\rangle=\int_{u=u_0} d\Sigma^v\ \hat{\phi}_W(X_2) \left(\hat{\phi}_V(X_{u_0}) \stackrel{\leftrightarrow}{\partial_v} G^R_V\left(X_{u_0},X_1\right)\right) |0\rangle.
\end{align}
Choosing a null surface $\Sigma_{v_0}$ of constant $v=v_0$ such that any $X_{v_0} \in \Sigma_{v_0}$ is either in the past of $X_1,X_3$ or spacelike-separated from them, and applying a similar construction to the operator $\phi_W(X_4)$, we end up with 
\begin{align}
\label{Psi}
|\Psi\rangle=\int_{v=v_0} d\Sigma^{u}\ \hat{\phi}_V(X_3) \left(\hat{\phi}_W(X_{v_0}) \stackrel{\leftrightarrow}{\partial_{u}} G^R_W\left(X_4,X_{v_0}\right)\right) |0\rangle.
\end{align}
Taking the overlap of \eqref{Psi'} and \eqref{Psi} then leads to an expression where only time-ordered four-point correlators appear:
\begin{align}
\label{overlap_general}
\langle \Psi'|\Psi \rangle&=\int_{v=v_0} d\Sigma^u \int_{u=u_0} d\Sigma^v \times\\
\nonumber
&\Big[\partial_v G^R_V\left(X_{u_0},X_1\right) \partial_u G^R_W\left(X_4,X_{v_0}\right) \langle \phi_V(X_{u_0}) \phi_W(X_2) \phi_V(X_3) \phi_W(X_{v_0})\rangle \\
\nonumber
&-\partial_v G^R_V\left(X_{u_0},X_1\right) G^R_W\left(X_4,X_{v_0}\right) \langle \phi_V(X_{u_0}) \phi_W(X_2) \phi_V(X_3) \partial_u\phi_W(X_{v_0})\rangle \\
\nonumber
&- G^R_V\left(X_{u_0},X_1\right) \partial_u G^R_W\left(X_4,X_{v_0}\right) \langle \partial_v\phi_V(X_{u_0}) \phi_W(X_2) \phi_V(X_3) \phi_W(X_{v_0})\rangle \\
\nonumber
&+ G^R_V\left(X_{u_0},X_1\right) G^R_W\left(X_4,X_{v_0}\right) \langle \partial_v\phi_V(X_{u_0}) \phi_W(X_2) \phi_V(X_3) \partial_u\phi_W(X_{v_0})\rangle\Big].
\end{align}

\subsubsection*{WKB and classical gravity limits}
The time-ordered correlation functions appearing in \eqref{overlap_general} may be represented as Feynman path integrals. For example, one can write    
\begin{subequations}
\label{path_integral}
\begin{align}
\nonumber
&\langle \phi_V(X_{u_0}) \phi_W(X_2) \phi_V(X_3) \partial_u\phi_W(X_{v_0})\rangle\\
&= \int Dg_{\mu\nu}\ D\phi_W D\phi_V\ \phi_V(X_{u_0}) \phi_W(X_2) \phi_V(X_3) \partial_u\phi_W(X_{v_0})\ e^{iS\left[\phi_W,\phi_V,g_{\mu\nu}\right]}\\
&=\int Dg_{\mu\nu}\ \partial_u\langle \mathcal{T} \phi_W(X_2) \phi_W(X_{v_0}) \rangle_{g_{\mu\nu}} \langle \mathcal{T} \phi_V(X_{u_0}) \phi_V(X_3) \rangle_{g_{\mu\nu}}\ e^{iS_{EH}\left[g_{\mu\nu}\right]}.
\end{align}
\end{subequations}
where $\mathcal{T}$ indicates time-ordering.  The correlators of $\phi_V$ and $\phi_W$ have factorized in this expression because we are taking the action to be
the Einstein-Hilbert action plus the Klein-Gordon actions of the scalar fields,
\begin{equation}
S\left[\phi_W,\phi_V,g_{\mu\nu}\right]=S_{KG}\left[\phi_W,g_{\mu\nu}\right]+S_{KG}\left[\phi_V,g_{\mu\nu}\right]+S_{EH}\left[g_{\mu\nu}\right] \, ,
\end{equation}
so that the fields $\phi_V$ and $\phi_W$ only interact through gravity. Note that we can safely drop the time-ordering operator $\mathcal{T}$ in \eqref{path_integral} due to the respective location of the points $X_2, X_{v_0}, X_{u_0}, X_3$. Under the assumption that the insertions points $X_1,..., X_4$ are sufficiently close to the conformal boundary, one can now evaluate these propagators in the WKB approximation. As reviewed in Appendix \ref{appendix:WKB}, the WKB approximation for a propagator in a background spacetime with metric $g_{\mu\nu}$ is, for $X$ in the future of $Y$,
\begin{align}
\label{WKBpropagator}
\langle \phi_V(X) \phi_V(Y) \rangle_{g_{\mu\nu}} \approx A(X,Y)\ e^{im_V S_V(X,Y)},
\end{align}
with 
\begin{equation}
S_V(X,Y)=- \int_{Y}^X d\tau.
\end{equation}
with $S_V$ being the proper time along the 
 timelike geodesic going from $Y$ to $X$.  Thus, $\partial_{X^\mu} S_V(X)=k_\mu(X)$ is the future-directed momentum of this geodesic at the point $X$. This phase is also the action for a point-like particle following the geodesic trajectory, and varying it with respect to the background metric $g_{\mu\nu}$ gives the associated stress-tensor which acts as a source for perturbations of the background metric.
 Plugging \eqref{WKBpropagator} back into the path integral \eqref{path_integral}, and taking the saddle point approximation in the classical gravity limit, we hence get  
\begin{align}
\nonumber
&\langle \phi_V(X_{u_0}) \phi_W(X_2) \phi_V(X_3) \partial_u\phi_W(X_{v_0})\rangle\\
&\approx -im_W k_u(X_{v_0}) A_W(X_2,X_{v_0}) A_V(X_{u_0},X_3) e^{iS_{\text{on-shell}}\left[g_{\mu\nu}\right]},
\end{align} 
with
\begin{equation}
\label{onshell action}
S_{\text{on-shell}}\left[g_{\mu\nu}\right]=S_{EH}+m_W S_W(X_2,X_{v_0})+m_V S_V(X_{u_0},X_3)\big|_{\text{on-shell}}.
\end{equation}
The action to be evaluated on-shell is the one of gravity sourced by two highly energetic (lightlike) geodesics. At leading order in an inverse energy expansion, the on-shell metric is
\begin{equation}
\label{metric solution}
g_{\mu\nu}=g^{(0)}_{\mu\nu}+h^W_{\mu\nu}+h^V_{\mu\nu}+\mathcal{O}(h^V h^W),
\end{equation}
where $g_{\mu\nu}^{(0)}$ is the original background metric satisfying prescribed boundary conditions and $h_{\mu\nu}\equiv h^V_{\mu\nu}+ h^W_{\mu\nu}$ describes the gravitational shock waves associated to the two lightlike geodesics. It should be noted that \eqref{metric solution} would be an exact solution of Einstein's equations if only one geodesic were considered \cite{Sfetsos:1994xa}. Here however, we neglect nonlinearities associated to the dynamical interaction of the shock waves with each other because these are higher order in the Newton constant. The on-shell action \eqref{onshell action} can be expanded to linear order in $h_{\mu\nu}$, giving \cite{Kabat:1992tb,Shenker:2014cwa}  
\begin{align}
S_{\text{on-shell}}\left[g_{\mu\nu}\right]=S_{\text{on-shell}}\left[g_{\mu\nu}^{(0)}\right]+S^h\left[h_{\mu\nu};g^{(0)}_{\mu\nu}\right]+\mathcal{O}(h^2),
\end{align}
with
\begin{equation}
\label{eikonal general}
S^h\left[h_{\mu\nu};g^{(0)}_{\mu\nu}\right]=\frac{1}{4} \int h_{\mu\nu} T^{\mu\nu}=\frac{1}{4} \int \left(h^V_{\mu\nu} T_W^{\mu\nu}+h^W_{\mu\nu} T_V^{\mu\nu}\right)\equiv \delta,
\end{equation}
and where $T^{\mu\nu}\equiv T_V^{\mu\nu}+T_W^{\mu\nu}$ is the stress-tensor associated to the two geodesics. The linear piece is also known as the \textit{eikonal phase shift} $\delta$, and will play a central role in the next sections. Hence, we have 
\begin{align}
&\langle \phi_V(X_{u_0}) \phi_W(X_2) \phi_V(X_3) \partial_u\phi_W(X_{v_0})\rangle \approx\\
\nonumber
&-im_W k_u(X_{v_0}) A_W(X_2,X_{v_0}) e^{im_W S_W(X_2,X_{v_0})}\ A_V(X_{u_0},X_3) e^{im_V S_V(X_{u_0},X_3)}\ e^{iS_{EH}\left[g^{(0)}_{\mu\nu}\right]}\ e^{i\delta}.
\end{align}
Note that the phase factor $e^{iS_{EH}\left[g^{(0)}_{\mu\nu}\right]}$ is common to all observables computed on this background geometry, and is therefore irrelevant. The overlap integral \eqref{overlap_general} is then
\begin{align}
\langle \Psi'|\Psi \rangle&=\int_{v_0} d\Sigma^u\ A_W(X_2,X_{v_0}) e^{i m_W S_W(X_2,X_{v_0})} \int_{u_0} d\Sigma^v\ A_V(X_{u_0},X_3) e^{i m_V S_V(X_{u_0},X_3)}\\
\nonumber
&\times \Big[\partial_v G^R_V\ \partial_u G^R_W+im_W k_u \partial_v G^R_V\ G^R_W-im_V k_v G^R_V\ \partial_u G^R_W+m_W m_V k_u k_v G^R_V\ G^R_W\Big]e^{i\delta}.
\end{align}
It is left to apply the WKB approximation to the retarded propagators. For $X$ in the future of $Y$, these are
\begin{equation}
\label{Gret}
G^R(X,Y)=2 \text{Im}\ \langle \phi(X) \phi(Y) \rangle\approx 2A(X,Y) \sin m S(X,Y),
\end{equation}
which leads to
\begin{align}
\nonumber
\label{overlap}
\langle \Psi'|\Psi \rangle
&=-4m_W m_V \int_{v_0} d\Sigma^u\ k_u A_W(X_2,X_{v_0}) A_W(X_4,X_{v_0}) e^{i m_W S_W(X_2,X_{v_0})} e^{-i m_W S_W(X_4,X_{v_0})}\\
&\hspace{0.5cm} \times \int_{u_0} d\Sigma^v\ k_v A_V(X_{u_0},X_3) A_V(X_{u_0},X_1) e^{i m_V S_V(X_{u_0},X_3)}e^{-i m_V S_V(X_{u_0},X_1)}\ e^{i\delta}.
\end{align}
This final formula is the WKB version of the overlap integral presented in \cite{Shenker:2014cwa}. It has the important advantage of being expressed in position-space and is therefore easily applicable to any background spacetime. Using \eqref{hat phi} and \eqref{Gret}, one can write similar expressions for the scalar propagators,
\begin{align}
\nonumber
&\langle \phi_V(X_1) \phi_V(X_3) \rangle\\
\label{propagator_1}
&=2m_V \int_{u_0} d\Sigma^v\ k_v A_V(X_{u_0},X_1) A_V(X_{u_0},X_3) e^{im_V S_V(X_{u_0},X_3)}e^{-im_V S_V(X_{u_0},X_1)},\\
\nonumber
&\langle \phi_W(X_2) \phi_W(X_4) \rangle\\
\label{propagator_2}
&=-2m_W \int_{v_0} d\Sigma^u\ k_u A_W(X_{v_0},X_2) A_W(X_{v_0},X_4) e^{im_W S_W(X_2,X_{v_0})}e^{-im_W S_W(X_4,X_{v_0})}.
\end{align}
Hence, the crucial ingredient that prevents the OTOC \eqref{overlap} from factorizing into the product of the two propagators (\ref{propagator_1}-\ref{propagator_2}) is the eikonal phase shift $e^{i\delta}$ which encodes the gravitational interaction of highly energetic field excitations. 

Again, as the radial location of the  insertion points of the bulk operators ($X_i$) are taken to the boundary of AdS space, a radial rescaling of  (\ref{OTOC}) computes the OTOC in the dual conformal field theory on the boundary \cite{Banks:1998dd,Balasubramanian:1998de,Balasubramanian:1998sn,Balasubramanian:1999ri}.   We will perform all calculations in a near-boundary limit of the insertion points. So, the dependence of the correlation functions on time and the transverse space also compute these dependences in the dual field theory.

\section{OTOC in BTZ-Vaidya}\label{section:geodesics}
We want to apply the general OTOC formula \eqref{overlap} to the three-dimensional BTZ-Vaidya background spacetime. 
For this, we need to compute geodesics whose endpoints are close to the conformal boundary together with the eikonal phase shift $\delta$ associated to the gravitational interactions of these geodesics.\\

The three-dimensional BTZ-Vaidya spacetime can be described in retarded coordinates $(v,r,x)$ whose associated metric is
\begin{equation}
ds^2=-\left(r^2-\theta(v_s-v)r_-^2-\theta(v-v_s)r_+^2\right)dv^2+2dvdr+r^2dx^2.
\end{equation} 
The surface $v=v_s$ corresponds to the location of a shell of dust falling in from the asymptotic boundary. This spacetime is also simply understood as a gluing of two planar BTZ black holes along the surface $v=v_s$, of respective horizon radii $r_-$ and $r_+$. This is made manifest by performing the change of coordinate $v=t+\frac{1}{2r_h} \ln \frac{r-r_h}{r+r_h}$,
\begin{equation}
\label{metric BTZ tr}
ds^2=-(r^2-r_h^2)dt^2+(r^2-r_h^2)^{-1} dr^2+r^2 dx^2=ds^2_{\text{BTZ}},
\end{equation}
where $r_h$ is identified with $r_-$ if $v<v_s$, or with $r_+$ if $v>v_s$. For simplicity, we refer to these two planar black hole regions as BTZ$_-$ and BTZ$_+$, respectively. See Fig.~\ref{fig:Penrose}. Note that we work in units where the AdS radius has been set to unity. Note also that unlike the retarded coordinate $v$, the time coordinate $t$ is not continuous across the shell except in the limit $r\to \infty$ corresponding to the location of the conformal boundary. We will also often make use of the advanced-retarded coordinates
\begin{align}
\label{tr to uv}
\begin{cases}
u=t-\frac{1}{2r_h} \ln \frac{r-r_h}{r+r_h},\\
v=t+\frac{1}{2r_h} \ln \frac{r-r_h}{r+r_h},
\end{cases} \qquad \Leftrightarrow \qquad
\begin{cases}
t=\frac{v+u}{2},\\
r=r_h \frac{1+e^{r_h(v-u)}}{1-e^{r_h(v-u)}},
\end{cases}
\end{align}
in which the BTZ metric takes the form
\begin{equation}
\label{metric BTZ uv}
ds^2_{\text{BTZ}}=-\left(r^2-r_h^2\right)dudv+r^2dx^2.
\end{equation}
In this gauge, $r=r(u,v)$ is a function of the advanced and retarded coordinates through \eqref{tr to uv}. We specify the four insertion points of the OTOC in Schwarzschild coordinates $(t,r,x)$:
\begin{align*}
r_i&=\varepsilon^{-1}, \qquad i=1,2,3,4,\\
t_1&=t_-, \qquad t_3=t_--\epsilon_-, \qquad x_1=x_3\equiv x_-,\\
t_2&=t_+, \qquad t_4=t_+-\epsilon_+, \qquad x_2=x_4\equiv x_+.
\end{align*}
Here, we consider $\varepsilon \ll 1$ as a perturbative parameter measuring how close to the conformal boundary the operators are inserted. As we will see, it will be directly related to the inverse energy of the bulk geodesics connecting these points. The role of $\epsilon_-, \epsilon_+$ is to regulate the UV divergences due to the insertion of operators at the same boundary points, $X_1\approx X_3$ and $X_2\approx X_4$. We will always consider the $\phi_V$ operators as being inserted in BTZ$_-$ ($t_-<v_s$), while we do not constrain the $\phi_W$ insertions.\\ 

Following Section \ref{section:OTOC formula}, we choose two null surfaces $\Sigma_{u_0}$ and $\Sigma_{v_0}$ and study the timelike geodesics connecting these surfaces to the insertion points $X_1, X_2, X_3, X_4$. 
The requirement that no point $X_{v_0} \in \Sigma_{v_0}$ can be in the causal future of the insertion points $X_1, X_3$ means that this surface necessarily lies in BTZ$_-$, and we can choose it to be any constant $v=v_0<t_-$ surface. Similarly, $\Sigma_{u_0}$ should be chosen such that no point $X_{u_0} \in \Sigma_{u_0}$ lies in the causal past of $X_2, X_4$. This is achieved by choosing a surface of some constant $u=u_0$ in BTZ$_-$ and continuing it across the shell in BTZ$_+$. The precise value of $u_0$ does not actually matter, only the existence of such a surface. These surfaces are illustrated in Fig.~\ref{fig:Psi Vaidya}-\ref{fig:Psi' Vaidya}.\\

When working in the limit of small UV regulators $\epsilon_+, \epsilon_- \ll 1$, the WKB phases in the overlap formula $\eqref{overlap}$ simply differ by
\begin{align}
S_V(X_{u_0},X_3)&=S_V(X_{u_0},X_1)+\epsilon_-\ k_t(X_1)+\mathcal{O}(\epsilon_-^2),\\
S_W(X_4,X_{v_0})&=S_W(X_2,X_{v_0})-\epsilon_+\ k_t(X_2)+\mathcal{O}(\epsilon_+^2),
\end{align}
such that
\begin{align}
\label{phaseV}
e^{i m_V S_V(X_{u_0},X_3)}e^{-i m_V S_V(X_{u_0},X_1)}&=e^{im_V \epsilon_- k_t(X_1)},\\
\label{phaseW}
e^{i m_W S_W(X_2,X_{v_0})} e^{-i m_W S_W(X_4,X_{v_0})}&=e^{im_W \epsilon_+ k_t(X_2)}.
\end{align}
As reviewed in Appendix~\ref{appendix:geodesics}, the momenta $k_t(X_1)$ and $k_t(X_2)$ of the ingoing and outgoing geodesics are directly related to their conserved energies,\footnote{If $X_2$ lies in BTZ$_+$, the outgoing geodesic crosses the shell and a jump in energy occurs at the crossing point. In that case, we need to distinguish between the energy $e_{out}^-$ of the BTZ$_-$ segment of the geodesic and the energy $e_{out}^+=-k_t(X_2)$ of the BTZ$_+$ segment of the geodesic.}
\begin{equation}
k_t(X_1)=-e_{in}, \qquad k_t(X_2)=-e_{out},
\end{equation}
and depend implicitly on the geodesic endpoints $X_1, X_{u_0}$ and $X_2, X_{v_0}$. Hence, the overlap \eqref{overlap} and the coincident two-point functions (\ref{propagator_1}-\ref{propagator_2}) simplify to
\begin{align}
\langle \Psi'|\Psi \rangle
\nonumber
\label{overlap Vaidya}
&=-4m_W m_V \int_{v_0} d\Sigma^u\ k_u(X_{v_0}) A_W(X_2,X_{v_0})^2\ e^{-im_W \epsilon_+ e_{out}}\\
&\hspace{19mm}\times \int_{u_0} d\Sigma^v\ k_v(X_{u_0}) A_V(X_{u_0},X_1)^2\ e^{-im_V \epsilon_- e_{in}}\ e^{i\delta},\\
\label{propagator Vaidya 1}
\langle \phi_V \phi_V \rangle&=2m_V \int_{u_0} d\Sigma^v\ k_v(X_{u_0}) A_V(X_{u_0},X_1)^2\ e^{-im_V \epsilon_- e_{in}},\\
\label{propagator Vaidya 2}
\langle \phi_W \phi_W \rangle&=-2m_W \int_{v_0} d\Sigma^u\ k_u(X_{v_0}) A_W(X_2,X_{v_0})^2\ e^{-im_W \epsilon_+ e_{out}}.
\end{align}
These integrals over the points $X_{u_0}=(u_0,v,x) \in \Sigma_{u_0}$ and $X_{v_0}=(u,v_0,x') \in \Sigma_{v_0}$ are too difficult to be performed exactly. In the spirit of \cite{Shenker:2014cwa}, we perform these by stationary phase approximation in the regime of large field masses\footnote{It is natural to consider $\varepsilon$ and $\epsilon_-, \epsilon_+$ on the same footing since they all act as UV regulators. In that case, the stationary phase approximation can be performed within the $m_V, m_W\gg 1$ WKB regime.} $m_V \epsilon_- \gg \varepsilon$ and $m_W \epsilon_+ \gg \varepsilon$ (in units where the AdS radius is set to unity). We will show that the functional dependence of $e_{in}$ and $e_{out}$ on the integration variables $x, x',u,v$ are such that the saddle point of the integrands corresponds to a single pair of highly energetic radial geodesics. In that case, the normalized OTOC takes a particularly simple form,
\begin{align}
\label{normalised OTOC}
\frac{\langle \phi_V(t_-,x_-) \phi_W(t_+,x_+) \phi_V(t_-,x_-) \phi_W(t_+,x_+) \rangle }{\langle \phi_V \phi_V \rangle \langle \phi_W \phi_W \rangle}\approx e^{i\delta}\Big|_{\text{saddle}},
\end{align}
where it is implicit that all insertions happen close to the conformal boundary, $r_i=\varepsilon^{-1}$. Hence, the normalized OTOC equals the eikonal phase shift associated to the gravitational interaction of this single pair of highly energetic radial geodesics. In order to determine which pair of geodesics, we have to find the saddle point of $e_{in}$ and $e_{out}$. This is done in Sections~\ref{section:ingoing}-\ref{section:outgoing}. Having determined the dominant pair of geodesics, we compute the eikonal phase shift associated to their gravitational interaction in Section~\ref{section:eikonal}.\\ 

We start by stating some useful general results regarding timelike geodesics in a planar BTZ black hole, whose derivation may be found in Appendix~\ref{appendix:geodesics}. It is known that any locally AdS space can be viewed as a patch of a hyperboloid embedded in a higher-dimensional flat space, and that timelike geodesics are circles on this hyperboloid. However, not any pair of timelike-separated points $(t_a,r_a,x_a)$ and $(t_b,r_b,x_b)$ with $t_b>t_a$ in Schwarzschild coordinates, may be joined by circles. This is only possible for those satisfying
\begin{equation}
\label{constraint}
-1<\cos L_{ab}<1,
\end{equation}
with
\begin{equation}
\label{cosL BTZ}
\cos L_{ab}=r_h^{-2}\left(r_a r_b \cosh r_h(x_b-x_a)-\sqrt{(r_a^2-r_h^2)(r_b^2-r_h^2)} \cosh r_h(t_b-t_a)\right).
\end{equation}
If the constraint \eqref{constraint} is satisfied, $L_{ab}$ is the length of the geodesic connecting these points. In this case, the conserved energy and transverse momenta associated to the Killing vectors $\partial_t$ and $\partial_x$ are
\begin{align}
\label{e BTZ}
e&\equiv-\left(\partial_t\right)^\mu \dot{x}_\mu=\frac{\sqrt{(r_a^2-r_h^2)(r_b^2-r_h^2)}\sinh r_h(t_b-t_a)}{r_h \sin L_{ab}},\\
\label{j BTZ}
j&\equiv \left(\partial_x\right)^\mu \dot{x}_\mu=\frac{r_a r_b \sinh r_h(x_b-x_a)}{r_h \sin L_{ab}},
\end{align}
where the dot refers to the derivative with respect to the geodesic proper time. In addition, the $v$ component of the momentum of an ingoing geodesic and the $u$ component of the momentum of an outgoing geodesic are both equal to
\begin{align}
\label{kv ku}
k_v=k_u=-\frac{1}{2}\left(e+\sqrt{e^2-j^2-(r^2-r_h^2)+j^2r_h^2r^{-2}}\right),
\end{align}
where one should keep in mind that the value of $r$ varies along the geodesic. 

\subsection{Ingoing geodesics} \label{section:ingoing}
\begin{figure}[h!]
	\centering
	\includegraphics[scale=0.45]{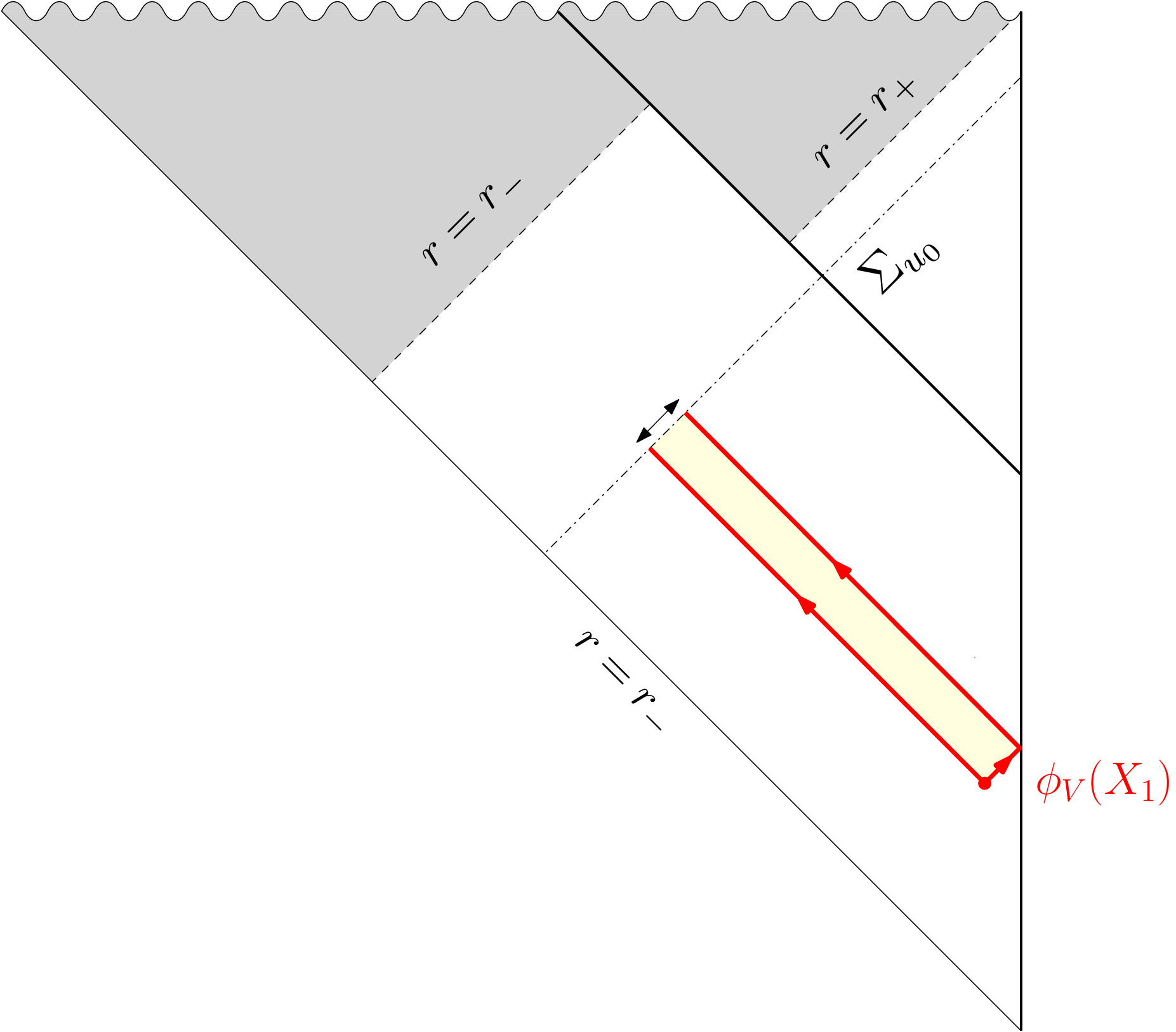}
	\caption{The region accessible to timelike geodesics connected to the insertion point $X_1$ is shown in yellow. It puts a restriction on the set of points $X_{u_0}=(u_0,v,x_1) \in \Sigma_{u_0}$ that may be connected by timelike geodesics to the insertion point $X_1=(t_1,\varepsilon^{-1},x_1)$ lying close to the conformal boundary.}
	\label{fig:geodesics}
\end{figure}

In order to find the saddle point of the energy $e_{in}$ along the null surface $\Sigma_{u_0}$, we consider those geodesics that connect the insertion point $X_1=(t_1,\varepsilon^{-1},x_1)$ to points $X_{u_0}=(u_0,v,x) \in \Sigma_{u_0}$ in BTZ$_-$. Assuming that such a geodesic exists, its energy and transverse momentum are found from (\ref{cosL BTZ}-\ref{j BTZ}),
\begin{align}
\label{e in}
e_{in}&=\frac{\sqrt{1-\varepsilon^2 r_-^2}}{\varepsilon \sin L_{in}}\frac{e^{r_-(v-t_1)}-e^{r_-(t_1-u_0)}}{1-e^{r_-(v-u_0)}},\\
\label{j in}
j_{in}&=\frac{\sinh r_-(x-x_1)}{\varepsilon \sin L_{in}} \frac{1+e^{r_-(v-u_0)}}{1-e^{r_-(v-u_0)}},
\end{align}
with
\begin{equation}
\label{cosL in}
\cos L_{in}=\frac{\left(1+e^{r_-(v-u_0)}\right) \cosh r_-(x-x_1)-\sqrt{1-\varepsilon^2r_-^2}\left(e^{r_-(t_1-u_0)}+e^{r_-(v- t_1)}\right)}{\varepsilon r_-\left(1-e^{r_-(v-u_0)}\right)}.
\end{equation}
As previously explained, the integrals over $x$ and $v$ in (\ref{overlap Vaidya}-\ref{propagator Vaidya 1}) are performed by stationary phase approximation. Since
\begin{equation}
\frac{\partial e_{in}}{\partial x}=\frac{\sqrt{1-\varepsilon^2 r_-^2}}{\varepsilon \sin^3 L_{in}}\frac{e^{r_-(v-t_1)}-e^{r_-(t_1-u_0)}}{1-e^{r_-(v-u_0)}}\cos L_{in}\frac{\partial\cos L_{in}}{\partial x},
\end{equation}
the $x$ integral localizes at values of $x$ satisfying $\cos L_{in}=0$ or $\frac{\partial \cos L_{in}}{\partial x} = 0$. Subsequently performing the $v$ integration, we similarly look for saddles along the $v$ direction. If $\cos L_{in}=0$, there is no saddle point along the $v$ direction provided that $u_0\neq t_1$. Hence, we consider the remaining case $\frac{\partial \cos L_{in}}{\partial x} = 0$, which is equivalent to $x=x_1$ and characterizes radial geodesics. Importantly, not every value of $v$ corresponds to a point $X_{u_0}=(u_0,v,x_1)$ that can be connected by a timelike geodesic to the insertion point $X_1$. From \eqref{constraint}, the existence of such a geodesic requires
\begin{equation}
\label{v constraint}
-\varepsilon <\frac{\left(1+e^{r_-(v-u_0)}\right)-\sqrt{1-\varepsilon^2r_-^2}\left(e^{r_-(t_1-u_0)}+e^{r_-(v- t_1)}\right)}{r_-\left(1-e^{r_-(v-u_0)}\right)}<\varepsilon.
\end{equation}
This is depicted in Fig.~\ref{fig:geodesics}. From (\ref{e in}-\ref{cosL in}), one can check that $e_{in}$ possesses a single saddle point satisfying the above restriction. To first order in $\varepsilon \ll 1$, it satisfies
\begin{align}
\label{v saddle}
v=\bar{v}+\mathcal{O}(\varepsilon^2), \qquad \bar{v} \equiv t_1.
\end{align}
In this limit, the timelike geodesic connecting $X_1$ to the unique saddle point $X_{u_0}=(u_0,t_1,x_1)$ follows a radial null ray of constant retarded coordinate given by \eqref{v saddle}, and the geodesic length can be shown to satisfy
\begin{equation}
\label{cos Lin saddle}
\cos L_{in}=\mathcal{O}(\varepsilon).
\end{equation}
From (\ref{e BTZ}-\ref{j in}), we also deduce its tangent momentum,
\begin{align}
\label{kv radial}
k_v=-e_{in}+\mathcal{O}(\varepsilon^0)=-\varepsilon^{-1}+\mathcal{O}(\varepsilon^0).
\end{align}

\subsection{Outgoing geodesics}\label{section:outgoing}
We similarly want to find the saddle point of the energy $e_{out}$ along the null surface $\Sigma_{v_0}$. For this, we need to distinguish between two types of outgoing geodesics, depending whether the insertion point $X_2=(t_2,\varepsilon^{-1},x_2)$ lies in BTZ$_-$ or in BTZ$_+$. In the former case, the treatment is parallel to the one of ingoing geodesics, such that the associated saddle lies at $x'=x_2,\ u=t_2+\mathcal{O}(\varepsilon)$. We therefore turn to geodesics that connect $X_2$ in BTZ$_+$ to points $X_{v_0}=(u,v_0,x')\in \Sigma_{v_0}$ in BTZ$_-$. 
Such a timelike geodesic necessarily crosses the shell at a point $X_s=(v_s,r_s,x_s)$. For that reason, we study segments of that geodesic on either side of the shell separately, and subsequently impose appropriate junction conditions at the shell that will fix the coordinates of the intermediate point $X_s$.       

\subsubsection*{BTZ$_+$ segment}
Assuming that it exists, the energy and transverse momentum of the timelike geodesic connecting a point $X_s$ on the shell to the insertion point $X_2$ are found from (\ref{cosL BTZ}-\ref{j BTZ}),
\begin{align}
\label{e out}
e_{out}^+&=\frac{\sqrt{1-\varepsilon^2 r_+^2}}{\varepsilon \sin L_{out}^+}\frac{r_s \sinh r_+(t_2-v_s)-r_+ \cosh r_+(t_2-v_s)}{r_+},\\
\label{j out}
j_{out}^+&=\frac{1}{\varepsilon \sin L_{out}^+}\frac{r_s \sinh r_+(x_2-x_s)}{r_+},
\end{align}
with
\begin{equation}
\label{cosL out}
\cos L_{out}^+=\frac{r_s \cosh r_+(x_2-x_s)-\sqrt{1-\varepsilon^2 r_+^2}\left[r_s \cosh r_+(t_2-v_s)-r_+\sinh r_+(t_2-v_s)\right]}{\varepsilon r_+^2}.
\end{equation}
The superscript `$+$' makes explicit that these quantities are associated to the BTZ$_+$ segment of the considered outgoing geodesic. Similarly, we will use the superscript `$-$' to denote quantities associated to the BTZ$_-$ segment of the outgoing geodesic. 

\subsubsection*{BTZ$_-$ segment}
We turn to the timelike geodesic segment in BTZ$_-$ connecting a point $X_{v_0}\in \Sigma_{v_0}$ to $X_s$. The energy and transverse momentum of the timelike geodesic are found from (\ref{cosL BTZ}-\ref{j BTZ}),
\begin{align}
\label{e minus}
e_{out}^-&=\frac{1}{\sin L_{out}^-} \frac{(r_s+r_-)e^{r_-(v_s-u)}-(r_s-r_-)e^{r_-(v_0-v_s)}}{1-e^{r_-(v_0-u)}},\\
j_{out}^-&=\frac{r_s \sinh r_-(x_s-x')}{\sin L_{out}^-} \frac{1+e^{r_-(v_0-u)}}{1-e^{r_-(v_0-u)}},
\end{align}
with
\begin{equation}
\label{cosL out minus}
\cos L_{out}^-=\frac{r_s\left(1+e^{r_-(v_0-u)}\right) \cosh r_-(x_s-x')-(r_s-r_-)e^{r_-(v_0-v_s)}-(r_s+r_-)e^{r_-(v_s-u)}}{r_-\left(1-e^{r_-(v_0-u)}\right)}.
\end{equation} 
Continuity of the retarded coordinates $(v,r,x)$ across the shell and a careful analysis of the geodesic equation in BTZ-Vaidya \cite{Aparicio:2011zy} imply that both segments of a BTZ-Vaidya geodesic have to satisfy the following junction conditions:
\begin{align}
\dot{x}_{out}^-\big|_{X_s}&=\dot{x}_{out}^-\big|_{X_s},\\
\label{conservation vdot}
\dot{v}_{out}^-\big|_{X_s}&=\dot{v}_{out}^+\big|_{X_s},
\end{align}
or equivalently \cite{Hubeny:2013dea}  
\begin{align}
\label{conservation j}
j_{out}^-&=j_{out}^+,\\
\label{conservation e}
e_{out}^-&=e_{out}^++\frac{r_+^2-r_-^2}{2}\ \dot{v}_{out}^+\big|_{X_s}.
\end{align}
These junction conditions may be shown to coincide with those originally derived in \cite{Balasubramanian:2011ur} by extremizing the total geodesic length $L_{out}=L_{out}^++ L_{out}^-$. They can be used to fix the coordinates $r_s, x_s$ of the intermediate point $X_s$ in terms of the endpoints $X_{v_0}$ and $X_2$, but are very difficult to solve in general. In Appendix~\ref{appendix:geodesics}, we show that the velocity of the geodesic at the shell is given by 
\begin{subequations}
\begin{align}
\dot{v}_{out}^+\big|_{X_s} &=\left(r_s^2-r_+^2\right)^{-1}\left(e_{out}^++\sqrt{(e_{out}^+)^2-(j_{out}^+)^2-(r_s^2-r_+^2)+r_+^2r_s^{-2}(j_{out}^+)^2}\right)\\
&=\frac{1}{r^2_+ \varepsilon \sin L^{+}_{out}} \left(  \sqrt{1-\varepsilon^2r_+^2} \cosh r_+\left(t_2 -v_s \right) - \cosh r_+ \left( x_s - x_2 \right) \right),
\label{vdot+ out}
\end{align}
\end{subequations}
and
\begin{equation}
\dot{v}_{out}^-\big|_{X_s} = \frac{ -(e^{r_-\left(v_0-v_s\right)} + e^{-r_-\left(u-v_s\right)})+ \left(1 + e^{r_-\left(v_0-u\right)}\right) \cosh r_-\left(x'-x_s \right)}{r_-\sin L^{-}_{out} \left(1-e^{r_-\left(v_0-u\right)}\right)} .
\label{vdot- out}
\end{equation}

\subsubsection*{Saddle point}
In order to perform the $x'$ integration in \eqref{overlap Vaidya} and \eqref{propagator Vaidya 2} by stationary phase approximation in the regime of large field masses, we need to find the associated saddle point of $e_{out}^+$. Here, we argue that a saddle point lies at $x'=x_2$ for any fixed value of $u$. From \eqref{e out}, one sees that $e_{out}^+$ depends on $x'$ only through $r_s(x')$ and $\cosh r_+(x_2-x_s(x'))$. Taking the square of the first junction condition \eqref{conservation j}, one can express $\cosh r_+(x_2-x_s(x'))$ in terms of $r_s(x')$ and $\cosh r_-(x_s(x')-x')$,
\begin{equation}
\{\cosh r_+(x_2-x_s(x'))\} \qquad \longrightarrow \qquad \{r_s(x'),\ \cosh r_-(x_s(x')-x') \}.
\end{equation}
The second junction condition \eqref{conservation e} allows to express $r_s(x')$ in terms of $\cosh r_-(x_s(x')-x')$,
\begin{equation}
\{ r_s(x') \} \qquad \longrightarrow \qquad \{ \cosh r_-(x_s(x')-x') \}.
\end{equation}
Hence, the energy $e_{out}^+$ associated to the BTZ$_+$ segment of the outgoing geodesic is shown to depend on the integration variable $x'$ only through the combination $\cosh r_-(x_s(x')-x')$, and a saddle point therefore exists for $x_s(x')=x'$. In turn, this implies that the associated outgoing geodesics have zero transverse momentum $j_{out}^-=j_{out}^+=0$ and that the saddle point lies at $x'=x_2$.\\ 

After having set $x'=x_2$, we look for saddles of $e_{out}^+$ along the $u$ direction. Only those intermediate points $X_s=(v_s,r_s(u),x_2)$ satisfying the constraint \eqref{constraint} may be connected by timelike geodesics to the insertion point $X_2$. To first order in $\varepsilon \ll 1$ and using \eqref{cosL out}, this constraint reads
\begin{equation}
\label{rs constraint}
-\frac{\varepsilon r_+^2}{1-\cosh r_+(t_2-v_s)}<r_s(u)-r_+ \coth \frac{r_+(t_2-v_s)}{2}<\frac{\varepsilon r_+^2}{1-\cosh r_+(t_2-v_s)}.
\end{equation}
Hence, only values of $u$ satisfying the above constraint should be considered when looking for saddle points of $e_{out}^+$. Taking the derivative of \eqref{e out}, we get
\begin{align}
\partial_u e_{out}^+&=\frac{\sqrt{1-\varepsilon^2 r_+^2}}{2\varepsilon^3 r_+^4 \sin^3 L_{out}^+}\ \partial_u r_s(u)\\
\nonumber
&\times \Big[\left(\sqrt{1-\varepsilon^2 r_+^2}\left(3+\cosh 2r_+(t_2-v_s)\right)-2\left(2-\varepsilon^2r_+^2\right) \cosh r_+(t_2-v_s)\right)r_s(u)\\
\nonumber
&\hskip 5mm -2r_+ \sinh r_+(t_2-v_s)\left(\sqrt{1-\varepsilon^2 r_+^2}\cosh r_+(t_2-v_s)-1\right)\Big]\  .
\end{align}
Hence, a saddle occurs when the term in square brackets vanishes. The corresponding value of $r_s(u)$ satisfies the constraint \eqref{rs constraint}, and is given to first order in $\varepsilon \ll 1$ by
\begin{equation}
\label{rs}
r_s(u)=r_+\coth \frac{r_+(t_2-v_s)}{2}+\mathcal{O}(\varepsilon^2) \qquad \left(\cos L_{out}^+=\mathcal{O}(\varepsilon)\right).
\end{equation} 
One can use the junction condition \eqref{conservation e} to determine the corresponding value of $u$. In the limit $\varepsilon \ll 1$, this problem considerably simplifies by noticing that the timelike geodesic connecting $X_2$ to the saddle point described above necessarily satisfies 
\begin{equation}
\label{cosL out saddle}
\sin L_{out}^-=\mathcal{O}(\varepsilon) \qquad \Leftrightarrow \qquad \cos L_{out}^- = 1 + \mathcal{O}(\varepsilon).
\end{equation}
From \eqref{cosL out minus} and (\ref{rs}-\ref{cosL out saddle}), one can therefore infer the associated value of $u$,
\begin{equation}
\label{u saddle}
u=\bar{u}+\mathcal{O}(\varepsilon), \qquad \bar{u}\equiv v_s-\frac{1}{r_-} \ln \frac{r_+ \sinh r_+(t_2-v_s)-r_- \cosh r_+(t_2-v_s)+r_-}{r_+ \sinh r_+(t_2-v_s)+r_- \cosh r_+(t_2-v_s)-r_-}.
\end{equation} 
In deriving the location of the saddle point, it has been implicitly assumed that the functions $r_s(u,x')$ and $x_s(u,x')$ were differentiable. A posteriori, we have explicitly checked that this was indeed the case, at least at the saddle point. Moreover, we numerically studied the dependence of $e^{+}_{out}$ on $x'$ and $u$, which suggested that the saddle we found is the only saddle.\\ 

As we pointed out, the timelike geodesic connecting $X_2$ to the saddle point described above follows a radial null ray in the small $\varepsilon$ limit. In particular, its BTZ$_-$ segment has constant advanced coordinate given by \eqref{u saddle}. Using \eqref{rs}, we may simplify the expressions of the conserved energy \eqref{e out} and initial velocity \eqref{vdot+ out} of the BTZ$_+$ geodesic segment,
\begin{align}
e_{out}^+&=\varepsilon^{-1}+\mathcal{O}\left(\varepsilon^0\right),\\
\dot{v}_{out}^+\big|_{X_s}&=\frac{\cosh r_+(t_2-v_s)-1}{r_+^2 \varepsilon}+\mathcal{O}(\varepsilon^0).
\end{align} 
Then, the junction condition \eqref{conservation e} allows us to evaluate the momentum \eqref{kv ku} of the BTZ$_-$ segment of the outgoing geodesic, 
\begin{subequations}
\label{ku radial}
\begin{align}
k_u&=-e_{out}^-+\mathcal{O}(\varepsilon^0)=-e_{out}^+-\frac{r_+^2-r_-^2}{2}\ \dot{v}_{out}^+\big|_{X_s}+\mathcal{O}(\varepsilon^0)\\
&=-\frac{\left(1-\frac{r_-^2}{r_+^2}\right)\cosh r_+(t_2-v_s)+1+\frac{r_-^2}{r_+^2}}{2\varepsilon}+\mathcal{O}\left(\varepsilon^0\right).
\end{align}
\end{subequations}

\subsection{Eikonal phase shift}\label{section:eikonal}
Having found the pair of highly energetic geodesics corresponding to the saddle point of (\ref{overlap Vaidya}-\ref{propagator Vaidya 2}), we are now in position to give the formula of the eikonal phase shift $\delta$ which encodes the interaction of one geodesic with the gravitational shock wave sourced by the other geodesic, and vice versa. When those interactions happen in BTZ$_-$, we show in Appendix~\ref{appendix:shocks} that the eikonal phase shift is given by
\begin{align}
\label{eikonal phase BTZ}
\delta=\frac{2\pi G_N m_V m_W}{r_-^2} k_v k_u\ e^{r_-(\bar{u}-\bar{v}-|x_+-x_-|)} \left(1-e^{-r_-(\bar{u}-\bar{v}-|x_+-x_-|)}\right)^2,
\end{align}
where $x_-, x_+, \bar{u}, \bar{v}$ label the position of the saddle point. If all insertions take place before the shell ($t_-,t_+ <v_s$), this interaction automatically happens in BTZ$_-$ and is nonzero only for
\begin{equation}
\left|\Delta x\right|<\Delta t, \qquad \Delta t\equiv t_+-t_-, \quad \Delta x\equiv x_+-x_-.
\end{equation}
Provided that this causality condition is satisfied, the eikonal phase shift is found from \eqref{eikonal phase BTZ} and (\ref{v saddle}-\ref{kv radial}),
\begin{equation}
\label{delta BTZ}
\delta=\frac{2\pi G_Nm_V m_W}{r_-^{2}\varepsilon^2}\ e^{r_-(\Delta t-|\Delta x|)} \left(1-e^{-r_-(\Delta t-|\Delta x|)}\right)^2, \qquad (t_-,t_+<v_s).
\end{equation}
If the second insertion takes place above the shell ($t_+>v_s$), the eikonal gravitational interaction may happen either in BTZ$_-$ or BTZ$_+$. In Appendix~\ref{appendix:shock Vaidya}, we show that the interaction happens in BTZ$_-$ if the transverse separation between the operators is sufficiently small:
\begin{equation}
\label{causality BTZVaidya}
\cosh r_- \Delta x<\frac{1+e^{-r_-(\bar{u}+v_s-2\bar{v})}}{e^{-r_-(v_s-\bar{v})}+e^{-r_-(\bar{u}-\bar{v})}}.
\end{equation}
The eikonal phase shift is then obtained from \eqref{eikonal phase BTZ}, \eqref{kv radial} and \eqref{ku radial},
\begin{align}
\label{delta BTZVaidya}
\delta&=\frac{\pi G_N m_V m_W}{r_-^2\varepsilon^2} \left(\left(1-\frac{r_-^2}{r_+^2}\right)\cosh r_+(t_+-v_s)+1+\frac{r_-^2}{r_+^2}\right)\\
\nonumber
&\times e^{r_-(\bar{u}-\bar{v}-|\Delta x|)} \left(1-e^{-r_-(\bar{u}-\bar{v}-|\Delta x|)}\right)^2, \qquad (t_-<v_s<t_+).
\end{align}
When \eqref{causality BTZVaidya} is not satisfied, one of the two gravitational interactions takes place in BTZ$_+$ and its computation goes beyond the scope of this paper. In particular, it requires to evolve the shock wave produced by the ingoing geodesic across the shell, i.e.~in the BTZ$_+$ region. For this one would have to solve the linearized Einstein's equations around the BTZ$_+$ background spacetime and subject to appropriate junction conditions at the shell.\\

As an interesting special case, we take the limit $r_- \to 0$ which corresponds to considering an AdS$_3$-Vaidya background instead of BTZ-Vaidya. The associated eikonal phase shift is worked out in Appendix~\ref{appendix:shock AdS} for the case where gravitational interaction occurs entirely in Poincaré AdS$_3$, and is given by
\begin{equation}
\label{eikonal phase AdS}
\delta=2\pi G_N\ m_V m_W\ k_v k_u \left(\bar{u}-\bar{v}-|\Delta x|\right)^2.
\end{equation}
Similarly to the BTZ-Vaidya case described above, two situations should be distinguished. If all insertions take place before the shell ($t_-,t_+ <v_s$), this interaction automatically happens in Poincaré AdS$_3$ and is nonzero only for
\begin{equation}
\left|\Delta x\right|<\Delta t.
\end{equation}
Provided that this causality condition is satisfied, the eikonal phase shift is found from \eqref{eikonal phase AdS} and \eqref{kv radial},
\begin{align}
\label{delta AdS}
\delta=\frac{2\pi G_N m_W m_V}{\varepsilon^2} \left(\Delta t-|\Delta x|\right)^2, \qquad (t_-,t_+<v_s).
\end{align}
If the second insertion takes place after the shell ($t_+>v_s$), the eikonal interaction does not necessarily happens in Poincaré AdS$_3$. In Appendix~\ref{appendix:shock Vaidya}, we show that it is only the case if 
\begin{equation}
\label{causality AdSVaidya}
\Delta x^2<(\bar{u}-\bar{v})(v_s-\bar{v}),
\end{equation}
where trajectories of the interacting radial geodesics follow constant advanced and retarded coordinates on the AdS$_3$ side, respectively,
\begin{equation}
\bar{u}=v_s+2 \frac{\cosh r_+(t_+-v_s)-1}{r_+ \sinh r_+(t_+-v_s)}, \qquad \bar{v}=t_-.
\end{equation}
The eikonal phase shift is then obtained from \eqref{eikonal phase AdS}, \eqref{kv radial} and \eqref{ku radial},
\begin{align}
\label{delta AdSVaidya}
\delta=\frac{\pi G_N m_W m_V}{\varepsilon^2} \left(\cosh r_+(t_+-v_s)+1\right) \left(\bar{u}-\bar{v}-|\Delta x|\right)^2, \qquad (t_-<v_s<t_+).
\end{align}
When \eqref{causality AdSVaidya} is not satisfied, one of the two eikonal interactions happens in BTZ$_+$ and we do not have the appropriate formula at our disposal, as in the more general BTZ-Vaidya case.  

\section{Results}\label{section:results}
We now discuss the results obtained in the WKB and stationary phase approximations, for the various cases studied in the previous sections. The expectation value of the commutator squared $D(t_+,x_+,t_-,x_-)$, which is the object of interest in the diagnosis of quantum chaos, is entirely controlled by the OTOC in the regime of time separation larger than the dissipation time scale, 
\begin{align}
\label{CS} 
D(t_+,x_+,t_-,x_-)&\approx 2\left(1- \frac{\text{Re}\ \langle \phi_V(t_-,x_-)\phi_W(t_+,x_+)\phi_V(t_-,x_-)\phi_W(t_+,x_+) \rangle}{\langle \phi_V \phi_V \rangle\ \langle \phi_W \phi_W \rangle}\right).
\end{align}
Two inverse `local' temperatures may be associated to a quenched state with BTZ-Vaidya dual,
\begin{equation}
\beta_-=\frac{2\pi}{r_-}, \qquad \text{and} \qquad \beta_+=\frac{2\pi}{r_+}.
\end{equation}
The inverse temperature $\beta_-$ is the one of the unperturbed thermal state before the quench is applied at $t=v_s$, while $\beta_+$ is the inverse temperature associated to the quenched state after it relaxed to equilibrium. The dissipation time scale is set by one of these two inverse temperatures in the following way. If all insertions happen before the quench ($t_-,t_+ <v_s$), then we are in the configuration described in \cite{Shenker:2014cwa} and the approximation \eqref{CS} requires $t_+-t_-\gg \beta_-$. If the first and second pairs of operators are inserted before and after the quench ($t_-<v_s<t_+$), respectively, then the above approximation applies in the regime $t_+-v_s \gg \beta_+$. In the previous sections, we derived the simple formula \eqref{normalised OTOC} for the OTOC in the limit of large scalar field masses $m_V, m_W \gg 1$ and to first order in a small $G_N$ expansion, such that
\begin{equation}
\label{CS_delta}
D(t_+,x_+,t_-,x_-)\approx \delta^2.
\end{equation} 
However, the WKB approximation has limitations that should be pointed out. Indeed, it corresponds to a high-frequency perturbative expansion for which propagating bulk fields only feel the local geometry of the background spacetime. Any physical effect associated to low-frequency or long-wavelength modes are nonperturbative from this perspective and are therefore missing from our description. In particular, this means that we cannot account for dissipation effects. On the one hand, these should manifest themselves for small time separations that coincide with the region of invalidity of \eqref{CS}. As explained in the Introduction, dissipation is also responsible for the saturation of the CS at very late times. The saturation behavior controlled by the quasinormal decay rate has been explicitly obtained in \cite{Shenker:2014cwa,Roberts:2014ifa} in case of a thermal state, but is missing from \eqref{CS_delta} due to the use of the WKB approximation. Hence, we should not trust the result for time separations such that~$\delta \gtrsim 1$.\\ 

\begin{figure}[h!]
	\centering
	\begin{subfigure}[b]{.49\textwidth}
		\centering
		\includegraphics[height=6.5cm]{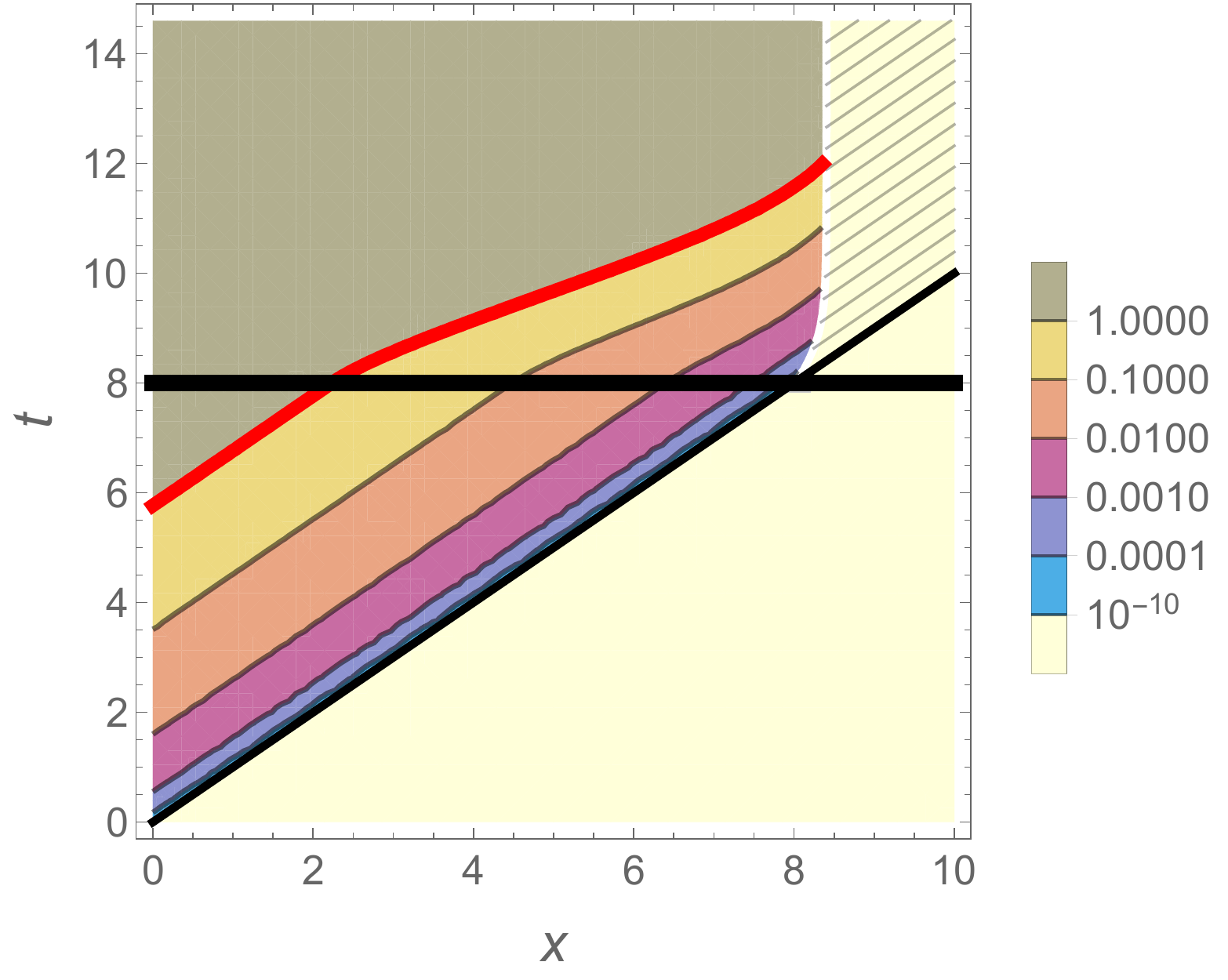}
		\caption{}
		\label{fig:BTZVaidya_3}
	\end{subfigure}\\
	\begin{subfigure}[b]{.49\textwidth}
		\centering
		\includegraphics[height=6.5cm]{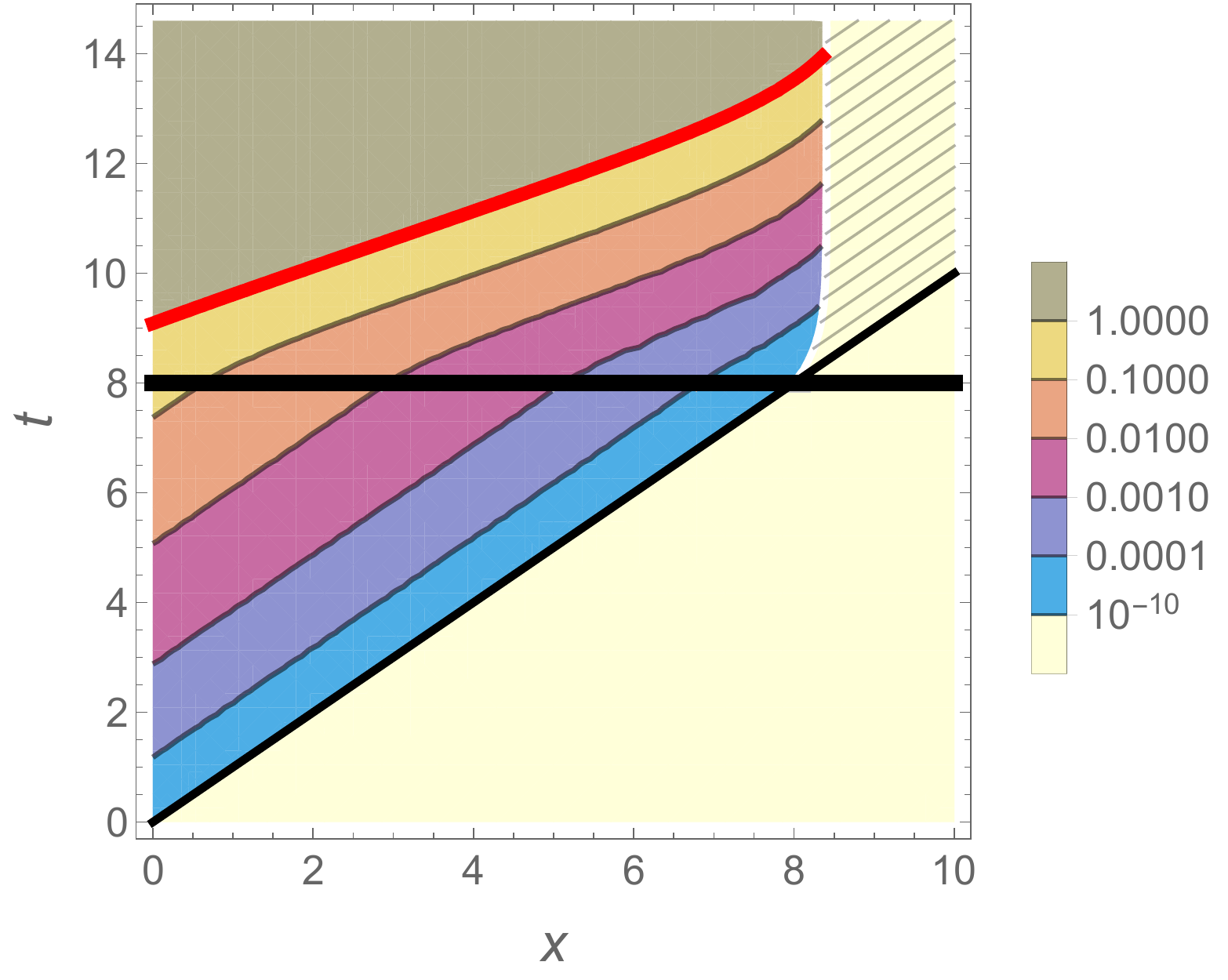}
		\caption{}
		\label{fig:BTZVaidya_1}
	\end{subfigure}
	\hskip 2mm
	\begin{subfigure}[b]{.49\textwidth}
		\centering
		\includegraphics[height=6.5cm]{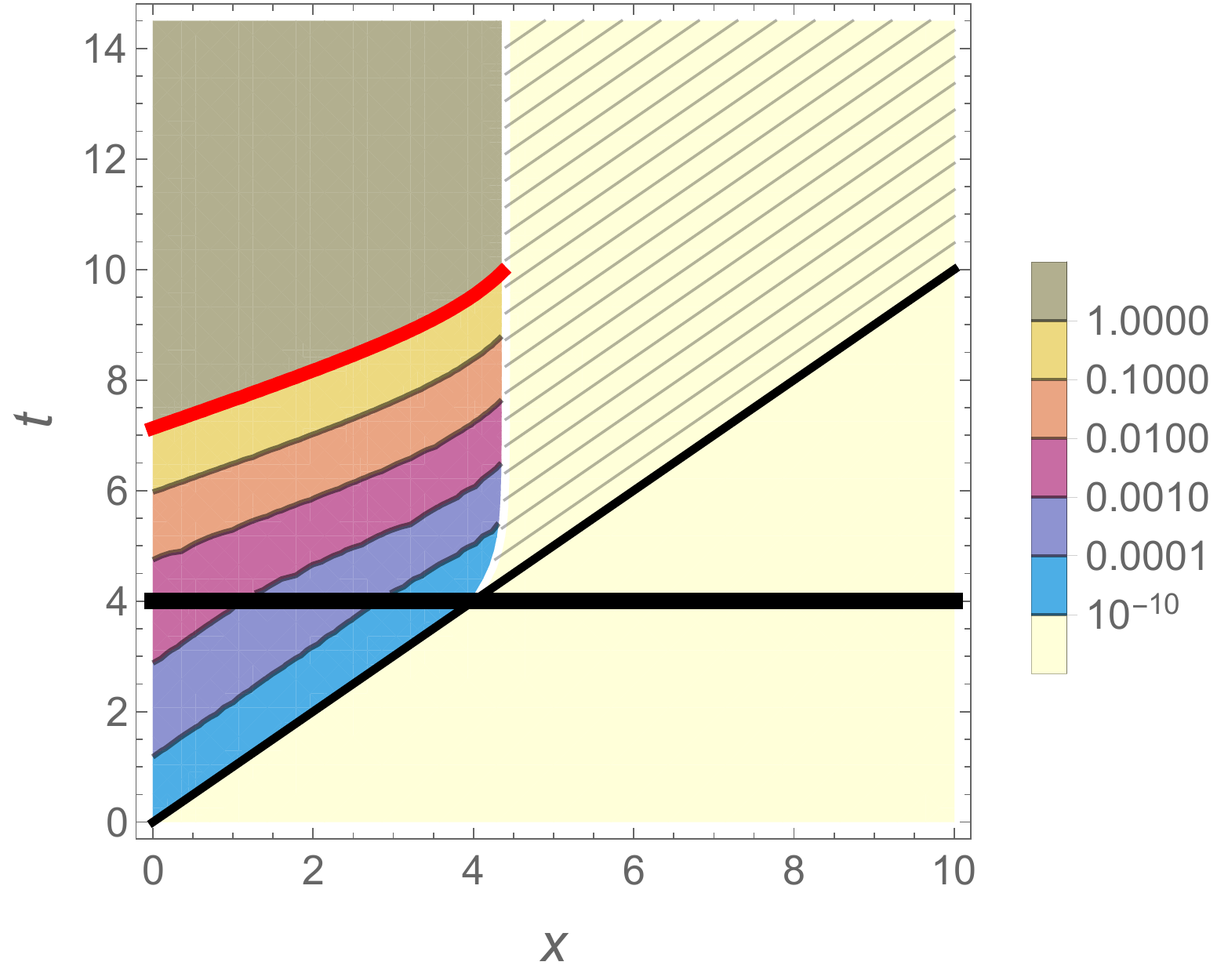}
		\caption{}
		\label{fig:BTZVaidya_2}
	\end{subfigure}
	\caption{Contour plots of the eikonal phase shift $\delta(t,x)$ in BTZ-Vaidya, for $m_V=m_W=1,\ \varepsilon=0.1,\ r_-=1,\ r_+=2$. The quench is applied at $t=v_s$ and is indicated by the black horizontal line, while the butterfly effect cone is indicated by the thick red line. The left endpoint of the red line, at $x=0$ and $t=t_*$, corresponds to efficient scrambling among local degrees of freedom. In the hatched regions, we do not have the appropriate formula of the eikonal phase shift at our disposal. (a) $G_N=5*10^{-6},\ v_s=8$. Fast scrambling is achieved before the quench is applied, and the butterfly velocity first coincides with the speed of light ($v_B=1$). After the quench, the butterfly velocity suddenly jumps to a superluminal value ($v_B=r_+/r_->1$) before gradually decreasing close to the hatched region. (b) $G_N=10^{-7},\ v_s=8$. (c) $G_N=10^{-7},\ v_s=4$. Fast scrambling is achieved after the quench is applied, such that the butterfly velocity takes a superluminal value from the onset ($v_B=r_+/r_->1$) before gradually decreasing close to the hatched region.}  
	\label{fig:BTZVaidya}
\end{figure}

The precise expression of the eikonal phase $\delta$ has been given in \eqref{delta BTZ}, \eqref{delta BTZVaidya} for the BTZ-Vaidya background, and in \eqref{delta AdS}, \eqref{delta AdSVaidya} for the AdS$_3$-Vaidya background as a particular limiting case. We use it to produce a series of contour plots in Figs.~\ref{fig:BTZVaidya}-\ref{fig:AdSVaidya} depicting the behavior of $\delta=\sqrt{D(t_+,x_+,t_-,x_-)}$ for different values of the shell insertion time $v_s$, indicated by the position of the black horizontal line. We set $t=t_+,\ x=x_+$ and $t_-=x_-=0$ without loss of generality.

\subsubsection*{BTZ-Vaidya}
Fig.~\ref{fig:BTZVaidya} shows the contour lines of $\delta(t,x)$ in BTZ-Vaidya, for two distinct positive values of the shell insertion time $v_s$. For $t<v_s$ corresponding to the lower parts of Figs.~\ref{fig:BTZVaidya_3}-\ref{fig:BTZVaidya_2}, contour lines approximately have unit slope. This can be simply understood by noting that for times larger than the dissipation time, equation \eqref{delta BTZ} leads to
\begin{equation}
\label{BTZ regime 1}
\delta(t,x)= \frac{2\pi G_Nm_V m_W}{\varepsilon^2 r_-^2}\ e^{r_-(t-|x|)}, \qquad (t<v_s).
\end{equation} 
This regime is therefore characterized by the maximal Lyapunov exponent allowed by a thermal state at inverse temperature $\beta_-$, namely $\lambda_-=r_-=2\pi \beta_-^{-1}$ \cite{Maldacena:2015waa}. To first order in a small $G_N$ expansion and in the regime $\delta \lesssim 1$, equation \eqref{BTZ regime 1} is in agreement with the results presented in \cite{Shenker:2014cwa,Roberts:2014ifa}. For $t>v_s$ corresponding to the upper parts of Figs.~\ref{fig:BTZVaidya_3}-\ref{fig:BTZVaidya_2}, new features appear. On the upper left side, one observes contour lines with approximately constant slope. This behavior is recovered from \eqref{delta BTZVaidya} together with \eqref{v saddle} and \eqref{u saddle} in the regime $r_+(t-v_s) \gg 1$ and $r_- (v_s-|\Delta x|) \gg 1$,
\begin{equation}
\label{BTZ regime 2}
\delta(t,x)=\frac{\pi G_N m_V m_W}{2\varepsilon^2}\ \frac{\left(r_++r_-\right)^2}{r_+^2 r_-^2} e^{r_- v_s} e^{r_+(t-v_s)} e^{-r_- |\Delta x|}, \qquad (t>v_s).
\end{equation}
Hence, in this regime we observe the maximal Lyapunov exponent associated to the second equilibrium temperature, $\lambda_+=r_+=2\pi \beta_+^{-1}$. From this equation, we see that a Lyapunov growth $e^{\lambda_- v_s}$ has been accumulated before the quench, and that the subsequent Lyapunov growth $e^{\lambda_+(t-v_s)}$ starts from $t=v_s$ onwards. The hatched region on the upper right sides of Figs.~\ref{fig:BTZVaidya_3}-\ref{fig:BTZVaidya_2} violates the condition \eqref{causality BTZVaidya}, and one would need to compute that part of the eikonal gravitational interaction happening in BTZ$_+$, in order to derive an expression appropriate to that region of the plot. Next to this hatched region, contour lines are bending upwards and we further comment on this behavior in Appendix~\ref{appendix:contour bending}. Finally, it can be noted that the transition between the regimes $t<v_s$ and $t>v_s$ is continuous. Indeed, we have
\begin{equation}
\lim\limits_{t\to v_s^+} \bar{u}=v_s=t, \qquad \lim\limits_{t\to v_s^+} k_u=-\varepsilon^{-1},
\end{equation} 
such that \eqref{delta BTZVaidya} smoothly transitions to \eqref{delta BTZ}.

As discussed in the Introduction, significant values of the commutator squared characterize regions where quantum information has been efficiently scrambled. The corresponding contour lines $(\delta=1)$ have been highlighted in red in Figs.~\ref{fig:BTZVaidya_3}-\ref{fig:BTZVaidya_2}. In this regard, two effects should be distinguished. The first one is the \textit{fast scrambling} among local degrees of freedom, while the second one is the spatial spreading of information. Depending on the values of the parameters, local scrambling is achieved either before or after the quench is applied. In the former case, it happens at the scrambling time
\begin{equation}
t_*\sim \lambda_-^{-1} \ln G_N^{-1}<v_s.
\end{equation}
Moreover, the boundary of the spatial region where information has propagated and has been significantly scrambled first grows ballistically at the speed of light ($v_B=1$) until the quench is applied, and subsequently \textit{faster} than the speed of light 
\begin{equation}
v_B=r_+/r_->1 \, .
\end{equation}
Hence, the butterfly effect cone ``opens up"; in other words chaos is superluminal. 

Eventually, the butterfly velocity $v_B$ decreases as contour lines are bending upwards, next to the hatched region. This situation is illustrated in Fig.~\ref{fig:BTZVaidya_3}. As a result, there is no conflict with causality since the commutator squared vanishes outside of the lightcone.   When local scrambling is achieved after the quench is applied, the scrambling time is given by
\begin{equation}
t_*\sim \left(1-\frac{\lambda_-}{\lambda_+}\right)v_s+\lambda_+^{-1}\ln G_N^{-1}>v_s.
\end{equation}
In this case, the boundary of the spatial region where information has propagated and has been significantly scrambled grows faster than the speed of light from the onset ($v_B=r_+/r_->1$) before eventually slowing down close to the hatched region. This situation is illustrated in Figs.~\ref{fig:BTZVaidya_1}-\ref{fig:BTZVaidya_2}. Again, there is no conflict with causality.

\subsubsection*{AdS$_3$-Vaidya}

\begin{figure}[t]
	\centering
	\begin{subfigure}[]{.49\textwidth}
		\centering
		\includegraphics[height=6.5cm]{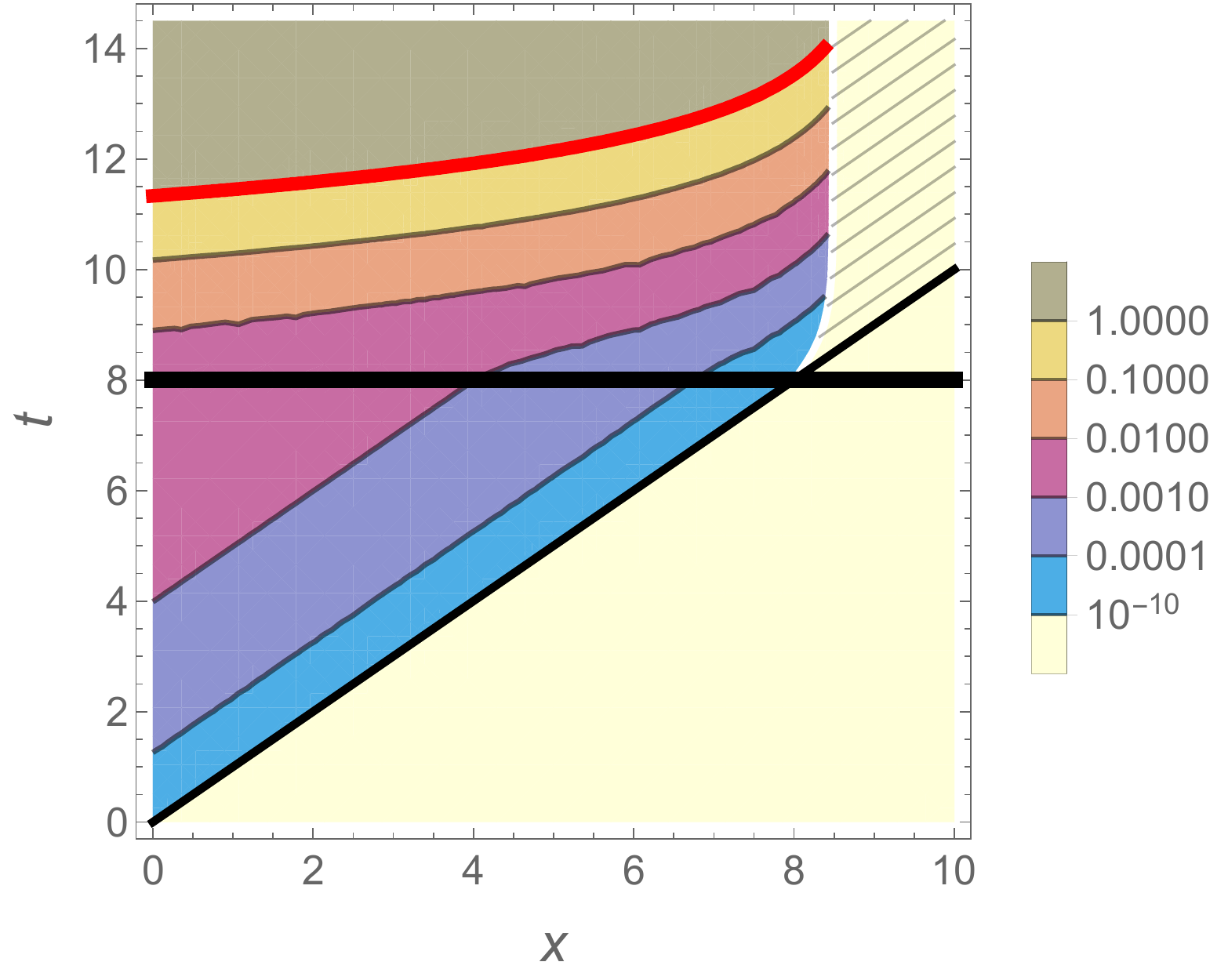}
		\caption{}
		\label{fig:AdSVaidya_1}
	\end{subfigure}
	\hskip 2mm
	\begin{subfigure}[]{.49\textwidth}
		\centering
		\includegraphics[height=6.5cm]{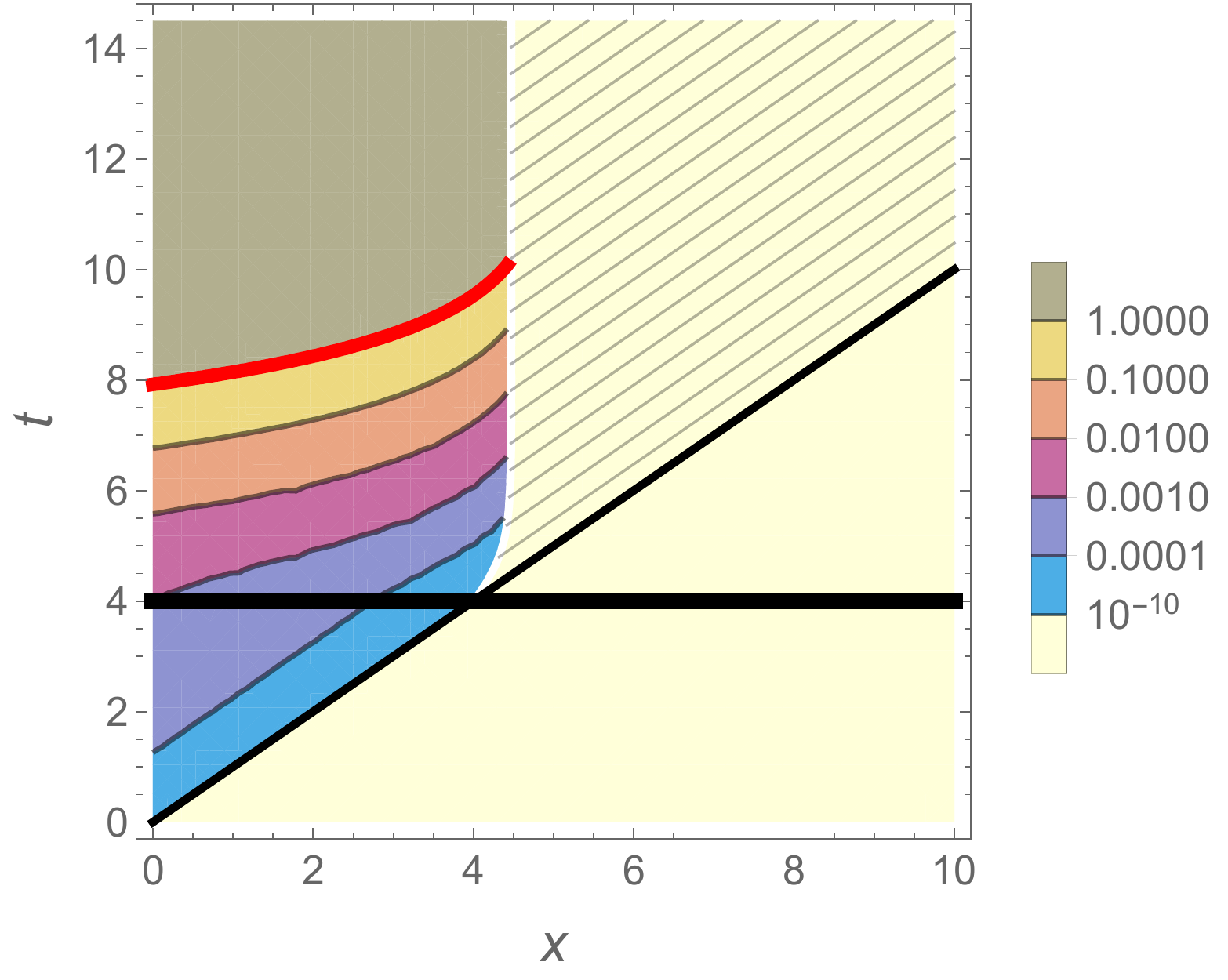}
		\caption{}
		\label{fig:AdSVaidya_2}
	\end{subfigure}
	\caption{Contour plots of the eikonal phase shift $\delta(t,x)$ in AdS$_3$-Vaidya, for $G_N=10^{-7},\ m_V=m_W=1,\ \varepsilon=0.1,\ r_-=0,\ r_+=1$. The quench is applied at $t=v_s$ and is indicated by the black horizontal line, while the butterfly effect cone is indicated by the thick red line. The left endpoint of the red line, at $x=0$ and $t=t_*$, corresponds to efficient scrambling among local degrees of freedom. In the hatched regions, we do not have the appropriate formula of the eikonal phase shift at our disposal. (a) $v_s=8$. (b) $v_s=4$. Fast scrambling is achieved after the quench is applied, such that the butterfly velocity takes a superluminal value from the onset and gradually decreases close to the hatched region. For $t>v_s$, the upward bending of the contour lines extends all the way from the left of the figures to the hatched regions, leaving no linear/ballistic regime.}
	\label{fig:AdSVaidya}
\end{figure}

Fig.~\ref{fig:AdSVaidya} shows the contour lines of $\delta(t,x)$ in AdS$_3$-Vaidya, for two distinct positive values of the shell insertion time $v_s$. For $t<v_s$ corresponding to the lower parts of Figs.~\ref{fig:AdSVaidya_1}-\ref{fig:AdSVaidya_2}, contour lines have exactly unit slope, as is manifest from \eqref{delta AdS}. In fact, this result agrees to first order in a small $G_N$ expansion with the one derived in \cite{Roberts:2014ifa} by CFT methods. In particular, the commutator growth is quadratic in $t-|x|$ rather than exponential, and has been qualified as \textit{slow scrambling} characteristic of vacuum in \cite{Roberts:2014ifa}. It should however be emphasized that the dissipation time is formally infinite in Poincaré AdS$_3$, such that the approximation \eqref{CS} is not reliable, and no direct relation between the eikonal phase $\delta$ and the commutator squared is expected. For $t>v_s$ corresponding to the upper parts of Figs.~\ref{fig:AdSVaidya_1}-\ref{fig:AdSVaidya_2}, the eikonal phase quickly behaves as
\begin{equation}
\label{AdS regime}
\delta(t,x)=\frac{\pi G_N m_W m_V}{2\varepsilon^2} e^{r_+(t-v_s)} \left(v_s+2r_+^{-1}-|x|\right)^2+\mathcal{O}\left(e^{-r_+ (t-v_s)}\right), \qquad (t>v_s).
\end{equation}
Hence, the commutator growth is controlled by the maximal Lyapunov exponent allowed by the late-time equilibrium temperature, $\lambda_+=2\pi \beta_+^{-1}$, and starts from $t=v_s$ onwards. As in BTZ-Vaidya, contour lines are bending upwards. This latter behavior extends all the way from the left of the figures to the hatched regions, leaving no linear/ballistic regime. Indeed, the time and space dependences in \eqref{AdS regime} do not occur through the combination $t-|x|$, such that there is no regime in which the spatial spreading is ballistic ($v_B\approx$ const), in contrast to the BTZ-Vaidya case. Neglecting the contribution from the initial period of slow scrambling, the scrambling time characterizing the successful scrambling of information among local degrees of freedom is simply given by
\begin{equation}
t_*\sim v_s + \lambda_+^{-1} \ln G_N^{-1}.
\end{equation} 
The boundary of the spatial region where information has propagated and has been significantly scrambled grows faster than the speed of light from the onset and gradually decreases without ever violating causality.  

\section{Conclusions and outlook}
In this paper, we computed the OTOC holographically in a state far from equilibrium, by investigating the chaotic behavior of a thermal state undergoing a sudden global injection of energy. The OTOC was constructed by considering the scattering of geodesics propagating in the BTZ-Vaidya spacetime, expressed in position space and using the WKB approximation. With this method, we recover the standard Lyapunov behavior in the case where the two pairs of operators are inserted before the quench. However, when one pair of operators is inserted before and the other pair after the quench, and the insertion locations are sufficiently close in space, the Lyapunov behavior saturates the bound set by the local temperature while the spatial suppression factor is set by the smallest temperature, causing the butterfly effect cones to become superluminal, albeit  in a way that preserves causality. 

There are several further directions that could be investigated. One next step would be to fill the unexplored (hatched) regions of Figs.~\ref{fig:BTZVaidya} and \ref{fig:AdSVaidya}. In the gravitational description, this would require solving Einstein's equations for the evolution of the shock wave created by an ingoing geodesic travelling in BTZ$_-$ past the infalling shell of matter. Another interesting question to ask is whether distinct quantities characterizing the spreading of information, such as entanglement and information velocities, also feature superluminal behavior. These velocities have been studied in a non-equilibrium context in \cite{Couch:2019zni}. 

One shortcoming of our method is that it does not describe the saturation behavior of the commutator squared, as the WKB approximation only accounts for high frequency modes. In \cite{Shenker:2014cwa} the states $|\Psi\rangle$ and $|\Psi' \rangle$ were decomposed in a momentum basis and the knowledge of the bulk-to-boundary propagators on the horizons made it possible to perform at least one of the four necessary integrals exactly. In this way, the Lyapunov behavior of the commutator squared as well as its saturation, where the lowest quasinormal mode dominates the integrals, were captured in a single formula. Inclusion of the saturation behavior in our setting would involve the bulk-to-boundary propagators in Vaidya spacetimes, which have been studied numerically in \cite{Keranen:2014lna,Keranen:2015mqc,David:2015xqa} for the case of AdS-Vaidya. 

It would also be interesting to understand how different aspects of our discussion, and especially the superluminal chaos, arise in the dual field theory picture. In \cite{Anous:2016kss,Anous:2017tza} a framework that allows for the computation of $n$-point functions in a quenched state starting from the vacuum was proposed, assuming domination of the Virasoro vacuum block exchange at large central charge.  Note that the field theory calculation of the OTOC in a thermal CFT ensemble \cite{Roberts:2014ifa}, performed using large central charge Virasoro block techniques, was sensitive to both the Lyapunov and the saturation regions. Therefore, formulating the computation of the OTOC in a quenched state in these terms might provide another way to tackle several of the questions raised by this paper.

\section*{Acknowledgments}
We thank Surbhi Khetrapal for related discussions. This work is supported in part by FWO-Vlaanderen through project G044016N and G006918N and by Vrije Universiteit Brussel through the Strategic Research Program ``High-Energy Physics.'' MDC is supported by a PhD fellowship from the Research Foundation Flanders (FWO).  VB is supported by the Simons Foundation through the It From Qubit Collaboration (Grant No. 38559), and also by the DOE (Contract No. FG02-05ER-41367).  VB thanks the Aspen Center for Physics (NSF Grant PHY-1607611) and the  Kavli Institute for the Physics and Mathematics of the Universe (IPMU, Tokyo) for hospitality as this work was completed.  

\appendix
\section*{Appendix}

\section{Timelike geodesics}\label{appendix:geodesics}
In this appendix, we collect useful information regarding timelike geodesics in planar BTZ~\cite{Cruz:1994ir}. We make use of the fact that planar BTZ is just a patch of pure AdS$_3$. The latter can be viewed as a hyperboloid in $4d$ flat space through the constraint
\begin{equation}
\label{hyperboloid}
\eta_{MN} X^M X^N=-1, \qquad \eta_{MN}=\text{diag}\left(-1,1,1,-1\right).
\end{equation}
We will refer to $X^M$ as `embedding coordinates'. Timelike geodesics can be shown to satisfy
\begin{align}
\label{geo_emb}
X^M(\tau)=P^M \cos\left(\tau-\tau_0\right)+Q^M \sin\left(\tau-\tau_0\right), \qquad P^2=Q^2=-1, \qquad P.Q=0,
\end{align}  
where $\tau$ is the geodesic proper time and $P^M, Q^M$ are integration constants. Given two points $X^M_a=X^M(\tau_a),\ X^M_b=X^M(\tau_b)$ along such a timelike geodesic, it directly follows that
\begin{equation}
\label{cosL}
\cos L_{ab} \equiv \cos \left(\tau_a-\tau_b\right)=-X_a.X_b.
\end{equation} 
This equation is very convenient to determine whether two points $X^M_a, X^M_b$ in pure AdS can be joined by a timelike geodesic. Indeed, such a geodesic exists if and only if the right-hand side of this equation lies in the interval $\left[-1,1\right]$. The integration constants are expressed in terms of the geodesic endpoints through 
\begin{equation}
P^M=X^M_a=X^M(\tau_0), \qquad Q^M=\frac{X^M_b-X^M_a \cos L_{ab}}{\sin L_{ab}}.
\end{equation}
The planar BTZ solution is obtained from flat embedding coordinates \eqref{hyperboloid} by the transformation
\begin{subequations}
\label{BTZ_coord}
\begin{align}
X^0&=\frac{\sqrt{r^2-r_h^2}}{r_h} \sinh r_h t,\\
X^1&=\frac{\sqrt{r^2-r_h^2}}{r_h} \cosh r_h t,\\
X^2&=\frac{r}{r_h} \sinh r_h x,\\
X^3&=\frac{r}{r_h} \cosh r_h x,
\end{align}
\end{subequations}
leading to the induced metric
\begin{equation}
ds_{BTZ}^2=-\left(r^2-r_h^2\right)dt^2+\left(r^2-r_h^2\right)^{-1}dr^2+r^2 dx^2.
\end{equation}
In planar BTZ, timelike geodesics have conserved energy $e$ and transverse momentum $j$ associated to the Killing vectors $\partial_t$ and $\partial_x$,
\begin{align}
\label{e}
e&\equiv -\left(\partial_t\right)^\mu \dot{x}_\mu=-g_{tt} \dot{t}=\left(r^2-r_h^2\right) \dot{t}>0,\\
j&\equiv \left(\partial_x\right)^\mu \dot{x}_\mu=g_{xx} \dot{x}=r^2 \dot{x},
\end{align}
such that the geodesic equation simply reads
\begin{equation}
\dot{r}^2=e^2-j^2-\left(r^2-r_h^2\right)+j^2r_h^2 r^{-2}.
\end{equation}
We can express the conserved quantities in terms of any two points $(t_a,r_a,x_a)$ and $(t_b,r_b,x_b)$ with $t_b>t_a$ along a geodesic, by using \eqref{geo_emb} together with the coordinate transformation \eqref{BTZ_coord},
\begin{align}
e&=\left(r^2-r_h^2\right) \frac{dt}{dX^M} \dot{X}^M=\frac{\sqrt{(r_a^2-r_h^2)(r_b^2-r_h^2)}\sinh r_h(t_b-t_a)}{r_h \sin L_{ab}},\\
j&=r^2 \frac{dx}{dX^M} \dot{X}^M=\frac{r_a r_b \sinh r_h(x_b-x_a)}{r_h \sin L_{ab}},\\
\cos L_{ab}&=r_h^{-2}\left(r_a r_b \cosh r_h(x_b-x_a)-\sqrt{(r_a^2-r_h^2)(r_b^2-r_h^2)} \cosh r_h(t_b-t_a)\right).
\end{align}
Note that the $M$ summation runs over any subset of three independent coordinates. In the main sections, we use the momentum of the geodesic in the $u-v$ directions defined in \eqref{metric BTZ uv}. For an ingoing geodesic, $\dot{r}<0$ such that
\begin{align}
\label{kv BTZ}
k_v&=g_{uv} \dot{u}=-\frac{1}{2}\left(e+\sqrt{e^2-j^2-(r^2-r_h^2)+j^2r_h^2r^{-2}}\right).
\end{align}
For an outgoing geodesic, $\dot{r}>0$ such that
\begin{align}
\label{ku BTZ}
k_u&=g_{uv} \dot{v}=-\frac{1}{2}\left(e+\sqrt{e^2-j^2-(r^2-r_h^2)+j^2r_h^2r^{-2}}\right).
\end{align}
We also need the velocity of an outgoing geodesic at its initial endpoint, 
\begin{align}
\dot{v}\big|_{r_a}&=\dot{t}+\left(r^2-r_h^2\right)^{-1}\dot{r}\big|_{r_a}=\left(r_a^2-r_h^2\right)^{-1}\left(e+\sqrt{e^2-j^2-(r_a^2-r_h^2)+r_h^2r_a^{-2}j^2}\right).
\end{align}

\section{WKB approximation} \label{appendix:WKB}
In this appendix, we describe the WKB approximation \cite{Stodolsky:1978ks,Hollowood:2008kq} to Klein-Gordon propagators satisfying 
\begin{equation}
\label{KG}
\square \phi(X)=m^2 \phi(X),
\end{equation}
assuming
\begin{equation}
\phi(X)=A(X)\ e^{im S(X)}, \qquad m\gg 1.
\end{equation}
The leading (quadratic) order in $m$ of \eqref{KG} leads to the so-called \textit{eikonal equation}
\begin{equation}
\qquad g^{\mu\nu}\partial_\mu S \partial_\nu S=-1,
\end{equation}
which may be written alternatively as
\begin{equation}
k^2=-1, \qquad k_\mu\equiv \partial_\mu S.
\end{equation}
Taking the covariant derivative of this equation and using $\nabla_{\mu} k_\nu=\nabla_{\nu} k_{\mu}$, it may be shown that $k^\mu$ satisfies the geodesic equation
\begin{equation}
k^\nu \nabla_\nu k_\mu=0,
\end{equation} 
and we conclude that the phase $S(X)$ is simply related to the action of a timelike geodesic ending at the point $X$,
\begin{equation}
S(X)=\int^X k_\mu dx^\mu=-\int^X d\tau.
\end{equation}
However, there is still freedom in choosing whether the tangent momentum $k^\mu$ is future- or past-directed. For the time-ordered Wightman propagator
\begin{equation}
\label{Wightman}
W(X,Y)=\langle \phi(X) \phi(Y) \rangle, \qquad  (X^0>Y^0),
\end{equation}
one has to choose $k_0<0$ or equivalently $k^0>0$, which corresponds to a geodesic traveling forward in time from $Y$ to $X$. As we use it in the main text, we also write the expression of the retarded propagator,
\begin{equation}
G_R(X,Y)=2 \text{Im}\ \langle \phi(X) \phi(Y) \rangle, \qquad (X^0>Y^0).
\end{equation}
At next-to-leading (linear) order in $m$, \eqref{KG} leads to
\begin{equation}
\frac{d}{d\tau}A=k^\mu \partial_\mu A=-\frac{A}{2} \theta, \qquad  \theta\equiv\nabla_\mu k^\mu.
\end{equation}
The solution for $A$ in $(d+1)$ spacetime dimensions is given in terms of the Van Vleck-Morette determinant $\Delta$ \cite{Visser:1992pz,Poisson:2003nc},
\begin{equation}
\label{WKBamplitude}
A(X)=\# \sqrt{\frac{\Delta(X)}{L(X)^d}},
\end{equation}
which satisfies
\begin{equation}
\frac{d}{d\tau} \Delta=\left(\frac{d}{L}-\theta\right)\Delta.
\end{equation}
Here, $L(X)=-S(X)$ is the length of the geodesic. In pure AdS$_{d+1}$, the Van Vleck-Morette determinant is given by \cite{Kent:2014nya}
\begin{equation}
\label{van vleck}
\Delta(X)=\left|\frac{L(X)}{\sin L(X)}\right|^d.
\end{equation}

\subsubsection*{Amplitude in BTZ-Vaidya and validity of the WKB approximation}
We now discuss the WKB amplitude of the Wightman propagator \eqref{Wightman} in the case of a BTZ-Vaidya background. Let us first consider $X, Y$ as being both on the BTZ$_-$ side of the shell. As we have seen, the amplitude may be expressed entirely in terms of the geodesic distance $L_-(X)$ between $X$ and some other reference point along the BTZ$_-$ segment of the geodesic, 
\begin{equation}
A(X)=C_0 \frac{\sqrt{\Delta\left[L_-(X)\right]}}{L_-(X)}.
\end{equation}
As the reference point can be chosen arbitrarily, the most general form of the solution is
\begin{equation}
A(X)=C_0 \frac{\sqrt{\Delta\left[L_-(X)-L_0\right]}}{L_-(X)-L_0}.
\end{equation}
Then, the fact that the source is located at $Y$ fixes the integration constant $L_0$,
\begin{equation}
L_0=L_-(Y).
\end{equation}
For example, we can conveniently choose 
\begin{equation}
L_-(X)=\int_{Y}^X d\tau, \qquad L_0=0.
\end{equation}
The other integration constant $C_0$ should be fixed by normalization of the amplitude.

We turn to the case where $X$ is lying in BTZ$_+$. The amplitude still assumes the same form,
\begin{equation}
A(X)=C_1 \frac{\sqrt{\Delta\left[L_+(X)-L_1\right]}}{L_+(X)-L_1},
\end{equation}   
where now $L_+(X)$ is the geodesic distance between $X$ and any other point along the BTZ$_+$ segment of the geodesic. One then uses continuity of the wave function at the shell in order to fix $L_1$ and $C_1$. This leads to
\begin{equation}
L_+(X)=\int_{X_{shell}}^{X} d\tau, \qquad L_1=\int^{Y}_{X_{shell}} d\tau, \qquad C_1=C_0.
\end{equation}
In the bulk of the paper, we do not explicitly need these formula. However, they are useful in assessing the validity of the WKB approximation. Indeed, it is assumed that variations of $A(X)$ are adiabatic with respect to the time scale $m^{-1}$,
\begin{equation}
\label{WKB validity}
\partial_X A(X)\ll m A(X).
\end{equation}
Looking at \eqref{WKBamplitude} and \eqref{van vleck}, we see that this breaks down for a short geodesic distance between $X$ and $Y$, i.e.\ when 
\begin{equation}
\int_Y^X d\tau \approx m^{-1}.
\end{equation}
The geodesics corresponding to the saddle points of Section~\ref{section:geodesics} have geodesic length $\pi/2$ up to $\mathcal{O}(\varepsilon)$ corrections, such that the WKB approximation for the associated propagators is valid. In addition, we avoid the evaluation of propagators whose endpoints are connected by a geodesic traveling very close to the shell. In such a case, moving $X$ on the other side of the shell might completely change the nature of the connecting geodesic and therefore lead to a discontinuity in the amplitude \eqref{WKBamplitude}. This would of course lead to a breakdown of \eqref{WKB validity}. 

\section{Shock waves and eikonal phases in AdS$_3$}\label{appendix:shocks}
Shock wave solutions in pure AdS have been described in various dimensions and coordinate systems in \cite{Hotta:1992qy,Podolsky:1997ri,Horowitz:1999gf,Afkhami-Jeddi:2017rmx}. We provide an alternative derivation of shock waves in any locally AdS$_3$ spacetime by solving Einstein's equations sourced by a null particle, starting with its embedding in 4-dimensional flat space where isometries are most straightforwardly realized. This allows to make direct contact between null geodesics and their gravitational field. We then apply this general construction to the case of radial null geodesics in planar BTZ. The shock wave associated to radial null geodesics in Poincaré AdS$_3$ is obtained as a limiting case.\\

First, we recall how to obtain the stress-tensor of a point-like particle. The action of a massive particle following a trajectory $X^\mu(s)$ is given by
\begin{equation}
S\left[X(s)\right]=-m \int ds\ \sqrt{g_{\mu\nu}(X(s)) \frac{dX^\mu}{ds} \frac{dX^\nu}{ds}},
\end{equation} 
where $s$ is any parameter along the geodesic. The particle stress-tensor is then obtained by functional differentiation with respect to the metric,
\begin{align}
T^{\mu\nu}(x)&=-\frac{2}{\sqrt{-g}} \frac{\delta S}{\delta g_{\mu\nu}(x)}=\frac{m}{\sqrt{-g}} \int ds\ \frac{\dot{X}^\mu \dot{X}^\nu}{\sqrt{-g_{\mu\nu} \dot{X}^\mu \dot{X}^\nu}}\ \delta^3(x-X(s)).
\end{align}
We further choose a set of coordinates $x^\mu=(u,v,y)$ where $u,v$ are lightlike, such that $X^\mu(s)=(U(s),V(s),Y(s))$. Changing of integration variable $s\to U$ directly leads to
\begin{subequations}
\begin{align}
T^{\mu\nu}(u,v,y)&=\frac{m}{\sqrt{-g}} \frac{ds}{dU} \frac{\dot{X}^\mu \dot{X}^\nu}{\sqrt{-g_{\mu\nu}\dot{X}^\mu \dot{X}^\nu}}\ \delta(v-V(s))\delta(y-Y(s))\Big|_{U(s)=u}\\
&=\frac{m}{\sqrt{-g}} \frac{d\tau}{dU}\ \dot{X}^\mu \dot{X}^\nu\ \delta(v-V(\tau))\delta(y-Y(\tau))\Big|_{U(\tau)=u},
\end{align}
\end{subequations}
where we specialized to the proper time $s=\tau$ in the last line. In a limit where the particle follows a null ray at $V(\tau)=v_0,\ Y(\tau)=y_0$, the stress-tensor becomes
\begin{subequations}
\label{Tuu}
\begin{align}
T^{uu}&=m k^u\ \frac{\delta(v-v_0)\delta(y-y_0)}{\sqrt{-g}}, \qquad k^u\equiv \frac{dU(\tau)}{d\tau},\\
T^{\mu\nu}&=0, \quad \mu,\nu \neq u.
\end{align}
\end{subequations}
\newline
We now consider a locally AdS$_3$ spacetime, and work in flat embedding coordinates \eqref{hyperboloid}. We eliminate the coordinate $X^3$ such that the AdS metric becomes
\begin{equation}
ds^2=\eta_{\mu \nu} dX^\mu dX^\nu - \frac{\left(\eta_{\mu\nu}X^\mu dX^\nu\right)^2}{1+\eta_{\mu\nu}X^\mu X^\nu} \qquad \mu=0,1,2.
\end{equation}
We then define lightcone coordinates,
\begin{equation}
\mathcal{V}=X^0+X^1, \qquad \mathcal{U}=X^0-X^1, \qquad \mathcal{Z}=X^2,
\end{equation}
such that 
\begin{align}
ds_{AdS}^2&=-d\mathcal{\mathcal{U}}d\mathcal{\mathcal{V}}+d\mathcal{Z}^2-\frac{\left(\mathcal{U}d\mathcal{V}+\mathcal{V}d\mathcal{U}-2\mathcal{Z} d\mathcal{Z}\right)^2}{4\left(1-\mathcal{U}\mathcal{V}+\mathcal{Z}^2\right)}, \qquad 
\sqrt{-g}=\frac{1}{2\sqrt{1-\mathcal{U}\mathcal{V}+\mathcal{Z}^2}}.
\end{align}
We consider a particle following a null trajectory along $\mathcal{V}=0$ and $\mathcal{Z}=\mathcal{Z}_0$, such that its stress-tensor is given by
\begin{align}
T_{\mathcal{V}\mathcal{V}}=E \sqrt{1+\mathcal{Z}^2}\ \delta(\mathcal{V}) \delta(\mathcal{Z}-\mathcal{Z}_0), \qquad E\equiv-m k_\mathcal{V}.
\end{align} 
Note that the momentum $k_\mathcal{V}$ is constant over the trajectory. We then look for the associated shock wave geometry from the following ansatz,
\begin{equation}
\label{shock wave}
ds^2=ds^2_{AdS}+ds^2_{SW}, \qquad ds^2_{SW}=16\pi G_N E\ \Pi(\mathcal{Z})\delta(\mathcal{V})d\mathcal{V}^2.
\end{equation}  
The only nontrivial component of Einstein's equations which needs to be solved is, using $\mathcal{V}\delta(\mathcal{V})'=-\delta(\mathcal{V})$,
\begin{equation}
\left[\left(1+\mathcal{Z}^2\right)\partial_\mathcal{Z}^2+\mathcal{Z} \partial_\mathcal{Z}-1\right]\Pi(\mathcal{Z})=-\sqrt{1+\mathcal{Z}^2}\ \delta(\mathcal{Z}-\mathcal{Z}_0).
\end{equation}
This can be recognized as the equation for a scalar propagator of unit mass in the transverse hyperbolic space spanned by $\vec{X}=\left(X^2, X^3\right)$,
\begin{equation}
\left[\nabla^2_\perp -1\right] \Pi(\vec{X})=-\frac{\delta(\vec{X}-\vec{X}_0)}{\sqrt{g_\perp}}.
\end{equation}
This equation appeared in \cite{Cornalba:2006xk,Cornalba:2007zb}, and its solution is well-known to be\footnote{See equation (6.12) of the review \cite{DHoker:2002nbb}.} 
\begin{align}
\label{propagator}
\Pi(\vec{X})&=\frac{|\xi|}{4} {}_2F_1\left[1,\frac{1}{2};2;\xi^2\right]=\frac{1}{2}\frac{|\xi|}{1+\sqrt{1-\xi^2}},
\end{align}
where we have used equation 9.121.24 of \cite{Gradshteyn} for the last step, and where the two-point invariant in transverse hyperbolic space is
\begin{equation}
\qquad \xi^{-1}\equiv -\vec{X}.\vec{X}_0= X^2 X_0^2-X^3 X_0^3.
\end{equation}
The shock wave solution \eqref{shock wave} describes the gravitational field of a highly energetic particle in any locally AdS$_3$ spacetime, assuming that one knows the coordinate transformation that maps it to the flat embedding coordinates. In addition, one can make use of AdS isometries in order to map other null trajectories to the one described above.   

\subsection{BTZ} \label{appendix:shock BTZ}
We give the shock wave geometry associated to a radial null geodesic in planar BTZ whose trajectory is characterized by $V=\bar{V},\ x=x_{in}$ in Kruskal coordinates. We write the transformation between Kruskal coordinates and \textit{some} embedding coordinate system $\bar{X}^i$,
\begin{subequations}
\begin{align}
\bar{X}^0&=\frac{V+U}{1+UV},\\
\bar{X}^1&=\frac{V-U}{1+UV},\\
\bar{X}^2&=\frac{1-UV}{1+UV} \sinh r_h x,\\
\bar{X}^3&=\frac{1-UV}{1+UV} \cosh r_h x.
\end{align}
\end{subequations}
The following AdS isometry maps the particle trajectory to the null ray $\mathcal{V}=\mathcal{Z}=0$,
\begin{equation}
\begin{pmatrix}
X^0\\
X^1\\
X^2\\
X^3
\end{pmatrix}
=
\begin{pmatrix}
1 & 0 & 0 & 0\\
0 & \cosh a & 0 & \sinh a\\
0 & 0 & 1 & 0\\
0 & \sinh a & 0 & \cosh a\\
\end{pmatrix}
\begin{pmatrix}
1 & 0 & 0 & 0\\
0 & 1 & 0 & 0\\
0 & 0 & \cosh r_h x_{in} & -\sinh r_h x_{in}\\
0 & 0 & -\sinh r_h x_{in} & \cosh r_h x_{in}
\end{pmatrix}
\begin{pmatrix}
\bar{X}^0\\
\bar{X}^1\\
\bar{X}^2\\
\bar{X}^3
\end{pmatrix},
\end{equation}
where $a\equiv \ln \frac{1-\bar{V}}{1+\bar{V}}$. Hence, we have
\begin{subequations}
\begin{align}
\mathcal{V}&=X^1+X^0=2\ \frac{V-\bar{V}^2 U+\bar{V}(UV-1)\cosh r_h\Delta x}{(1+UV)(1-\bar{V}^2)}, \qquad \Delta x\equiv x-x_{in},\\
\mathcal{Z}&=X^2=\frac{1-UV}{1+UV} \sinh r_h \Delta x,\\
\xi^2&=\frac{1}{1+\mathcal{Z}^2}=\frac{1}{1+\left(\frac{1-UV}{1+UV}\right)^2\sinh^2 r_h\Delta x}.
\end{align}
\end{subequations}
From this coordinate transformation and the general solution \eqref{shock wave}, we find the $VV$ component of the shock wave,
\begin{align}
\label{shock wave BTZ}
h_{VV}&=-16\pi G_N mk_V\ \frac{1+U\bar{V}\cosh r_h \Delta x}{1+UV}\ \delta\left(V-\bar{V} \frac{U\bar{V}+\cosh r_h\Delta x}{1+U\bar{V}\cosh r_h\Delta x}\right) \theta\left(V-\bar{V}\right) \Pi(\xi^2),
\end{align}
where we identified the momentum of the geodesic as
\begin{equation}
mk_V=mk_\mathcal{V} \frac{\partial \mathcal{V}}{\partial V}\Big|_{V=\bar{V},x=x_{in}}=\frac{2mk_\mathcal{V}}{1-\bar{V}^2}=-\frac{2E}{1-\bar{V}^2}.
\end{equation}
Note that without the step function $\theta(V-\bar{V})$, \eqref{shock wave BTZ} would also contain the shock wave associated to the reflected geodesic at the conformal boundary, namely the outgoing geodesic with trajectory $U=-\bar{V}^{-1},\ x=x_{in}$. The other components of $h_{\mu\nu}$ are not required for what follows. For $\bar{V}=0$, we recover the shock wave solution discussed in \cite{Shenker:2014cwa},
\begin{align}
h_{VV}=-8\pi G_N\ mk_V \exp\left(-r_h|\Delta x|\right)\delta(V) \theta(V).
\end{align}
In addition, the shock wave produced by an outgoing particle following the $U=\bar{U},\ x=x_{out}$ trajectory is found by interchanging $U$ and $V$ in \eqref{shock wave BTZ}.

\subsubsection*{Eikonal phase shift}
The eikonal interaction of an ingoing particle at $V=\bar{V},\ x=x_{in}$ and an outgoing particle at $U=\bar{U},\ x=x_{out}$ may now be computed. This interaction will be nontrivial only if the outgoing particle crosses the shock wave of the ingoing particle, and vice versa. For this, there should exist a point $(\bar{U},V,x_{out})$ along the outgoing geodesic such that the argument of the delta function in \eqref{shock wave BTZ} is zero, i.e.~such that
\begin{equation}
\left(1+\bar{U}\bar{V} \cosh r_h\Delta x \right)V=\left(\bar{U}\bar{V}+\cosh r_h \Delta x\right)\bar{V}, \qquad V-\bar{V}>0.
\end{equation}
Since $V\in \left(0,-\bar{U}^{-1}\right)$ for such an outgoing geodesic, we immediately get $\bar{U}\bar{V}>-1$, and a nontrivial interaction requires
\begin{equation}
\label{interaction condition}
\bar{V} < \bar{V} \frac{\cosh r_h\Delta x+\bar{U}\bar{V}}{1+\bar{U}\bar{V}\cosh r_h\Delta x}<-\frac{1}{\bar{U}},
\end{equation}
or equivalently,
\begin{equation}
\label{causality Kruskal}
\cosh r_h \Delta x<-\frac{\bar{U}\bar{V}+(\bar{U}\bar{V})^{-1}}{2}.
\end{equation}
In this case, using \eqref{Tuu} and \eqref{shock wave BTZ}, one has
\begin{subequations}
\begin{align}
\nonumber
&\delta=S_{on-shell}\\
&=\frac{1}{4}\int d^3x\ \sqrt{-g}\left(h_{UU} T^{UU}+h_{VV}T^{VV}\right)\\
&=4\pi G_N\ m_{in} m_{out}\ k_V^{in} k_U^{out}\left(1+\bar{U}\bar{V}\cosh r_h \Delta x\right)\left(1+\bar{U}\bar{V}\ \frac{\bar{U}\bar{V}+\cosh r_h \Delta x}{1+\bar{U}\bar{V}\cosh r_h \Delta x}\right) \Pi(\xi^2)\\
&=2\pi G_N\ m_{in} m_{out}\ k_V^{in} k_U^{out}\ e^{-r_h|\Delta x|} \left(1+\bar{U}\bar{V}e^{r_h|\Delta x|}\right)^2.
\end{align}
\end{subequations}
For either $\bar{U}=0$ or $\bar{V}=0$, the eikonal phase shift discussed in \cite{Shenker:2014cwa} is recovered,
\begin{equation}
\delta=2\pi G_N\ m_{in} m_{out}\ k_V^{in} k_U^{out} e^{-r_h\left|\Delta x\right|}.
\end{equation}
In the main text, we work in advanced-retarded Schwarzschild coordinates $(u,v,x)$, related to Kruskal coordinates by
\begin{equation}
\label{Kruskal to uv}
U=-e^{-r_h u}, \qquad V=e^{r_h v}.
\end{equation}
In this coordinate system, the requirement \eqref{causality Kruskal} of nontrivial eikonal interaction simply becomes
\begin{equation}
\label{causality uv}
\left|\Delta x\right|<\bar{u}-\bar{v},
\end{equation} 
while the eikonal phase shift is
\begin{align}
\label{delta appendix BTZ}
\delta&=2\pi G_N r_h^{-2}\ m_{in} m_{out}\ k_v^{in} k_u^{out}\ e^{r_h(\bar{u}-\bar{v}-|\Delta x|)}\ \left(1-e^{-r_h(\bar{u}-\bar{v}-|\Delta x|)}\right)^2.
\end{align}

\subsection{Poincaré AdS$_3$}\label{appendix:shock AdS}
The shock wave geometry associated to an ingoing radial null geodesic in Poincaré AdS$_3$ with trajectory characterized by $v=\bar{v},\ x=x_{in}$, can be obtained from \eqref{shock wave BTZ} in the limit $r_h \rightarrow 0$. Before taking this limit, one first needs to change to advanced-retarded Schwarzschild coordinates $(u,v,x)$ defined in \eqref{Kruskal to uv}, such that the $vv$ component of the shock wave is
\begin{align}
\label{shockwave AdS}
h_{vv}&=-16\pi G_N\ mk_v\ e^{r_h(v-\bar{v})}\ \frac{1-e^{r_h(\bar{v}-u)}\cosh r_h \Delta x}{1-e^{r_h(v-u)}}\\
\nonumber
&\times \delta\left(v-\frac{1}{r_h}\ln\left( e^{r_h \bar{v}} \frac{-e^{r_h(\bar{v}-u)}+\cosh r_h\Delta x}{1-e^{r_h(\bar{v}-u)}\cosh r_h\Delta x}\right)\right) \theta(v-\bar{v})\ \Pi(\xi^2),
\end{align}
with
\begin{equation}
\xi^2 = \frac{1}{1+\left( \frac{1+e^{r_h(v-u)}}{1-e^{r_h(v-u)}} \right)^2 \sinh^2 r_h\Delta x} .
\end{equation}
The limit $r_h \to 0$ may now be taken, resulting in
\begin{align}
h_{vv}&=-16\pi G_N\ mk_v\ \frac{u-\bar{v}}{u-v}\ \delta\left(v-\bar{v}-\frac{\Delta x^2}{u-\bar{v}}\right)\theta(v-\bar{v})\ \Pi(\xi^2),\\
\xi^2&=\frac{(u-v)^2}{(u-v)^2+4 \Delta x^2},
\end{align}
where evaluation of the transverse propagator gives
\begin{align}
\Pi(\xi^2)&=\frac{1}{2} \frac{u-\bar{v}-|\Delta x|}{u-\bar{v} + |\Delta x|}.
\end{align}
Putting these together, we find the following shock wave component  
\begin{equation}
\label{shock wave Poincare}
h_{vv}=-8\pi G_N\ mk_v\ \frac{\left(u-\bar{v}\right)^2}{\left( u-\bar{v}+|\Delta x|\right)^2} \ \delta\left(v-\bar{v}-\frac{\Delta x^2}{u-\bar{v}}\right)\theta(v-\bar{v}).
\end{equation}
This result agrees with the shock wave solution associated to a radial geodesic in Poincaré AdS$_3$ presented in \cite{Afkhami-Jeddi:2017rmx}.

\subsubsection*{Eikonal phase}
The eikonal interaction of an ingoing particle at $v=\bar{v},\ x=x_{in}$ and an outgoing particle at $u=\bar{u},\ x=x_{out}$ may now be computed. This interaction will be nontrivial only if the outgoing particle crosses the shock wave of the ingoing particle, and vice versa. For this, there should exist a point $(\bar{u},v,x_{out})$ along the outgoing geodesic such that the argument of the delta function in \eqref{shock wave Poincare} is zero, i.e.~such that
\begin{equation}
(\bar{u}-\bar{v})(v-\bar{v})-\Delta x^2=0, \qquad v-\bar{v}>0.
\end{equation}
Since $v\in \left(-\infty,\bar{u}\right)$ for such an outgoing geodesic, a nontrivial interaction again imposes the condition \eqref{causality uv}. If it is fulfilled, one has 
\begin{subequations}
\label{delta appendix AdS}
\begin{align}
\delta&=S_{on-shell}=\frac{1}{4}\int d^3x\ \sqrt{-g}\left(h_{uu} T^{uu}+h_{vv}T^{vv}\right)\\
&=2\pi G_N\ m_{in} m_{out}\ k_v^{in} k_u^{out} \left(\bar{u}-\bar{v}-|x_{in}-x_{out}|\right)^2.
\end{align}
\end{subequations}
This could have been obtained directly from \eqref{delta appendix BTZ} in the limit $r_h \to 0$.

\subsection{BTZ-Vaidya and AdS$_3$-Vaidya}\label{appendix:shock Vaidya}
The shock wave geometry associated to an ingoing radial null geodesic in BTZ$_-$ with trajectory characterized by $v=\bar{v}<v_s,\ x=x_{in}$, is given below the shell by the solution \eqref{shock wave BTZ}. However, part of the shock wave reaches the shell at $v=v_s$, at which point the solution \eqref{shock wave BTZ} is not valid anymore. Instead, one would have to evolve the shock wave solution across the shell through Einstein's equations, subject to appropriate junction conditions at the shell. This goes beyond the scope of this paper. Hence, with the shock wave solution in BTZ$_-$, we are only able to compute the eikonal phase shift associated to the gravitational interaction of this ingoing geodesic with an outgoing radial geodesic following the $u=\bar{u},\ x=x_{out}$ trajectory in case that the latter interacts with the shock wave before crossing the shell. The condition for this to happen requires a slight modification of \eqref{interaction condition}, namely 
\begin{equation}
\bar{V} < \bar{V} \frac{\cosh r_-\Delta x+\bar{U}\bar{V}}{1+\bar{U}\bar{V}\cosh r_-\Delta x}<e^{r_- v_s},
\end{equation}
or equivalently,
\begin{equation}
\cosh r_- \Delta x<\frac{1+e^{-r_-(\bar{u}+v_s-2\bar{v})}}{e^{-r_-(v_s-\bar{v})}+e^{-r_-(\bar{u}-\bar{v})}}.
\end{equation}
This condition is of course stronger than \eqref{interaction condition} or \eqref{causality uv}. If it is satisfied, the eikonal phase shift is still given by \eqref{delta appendix BTZ}. If not, we do not have the appropriate formula at our disposal. The same comments apply to the computation of the eikonal phase shift in AdS$_3$-Vaidya. In particular, it agrees with the Poincaré AdS$_3$ result \eqref{delta appendix AdS} if the following condition is satisfied,
\begin{equation}
\Delta x^2<(\bar{u}-\bar{v})(v_s-\bar{v}).
\end{equation}  

\subsection{Upward bending of contour lines}\label{appendix:contour bending}

\begin{figure}[t]
	\centering
	\includegraphics[height=6.5cm]{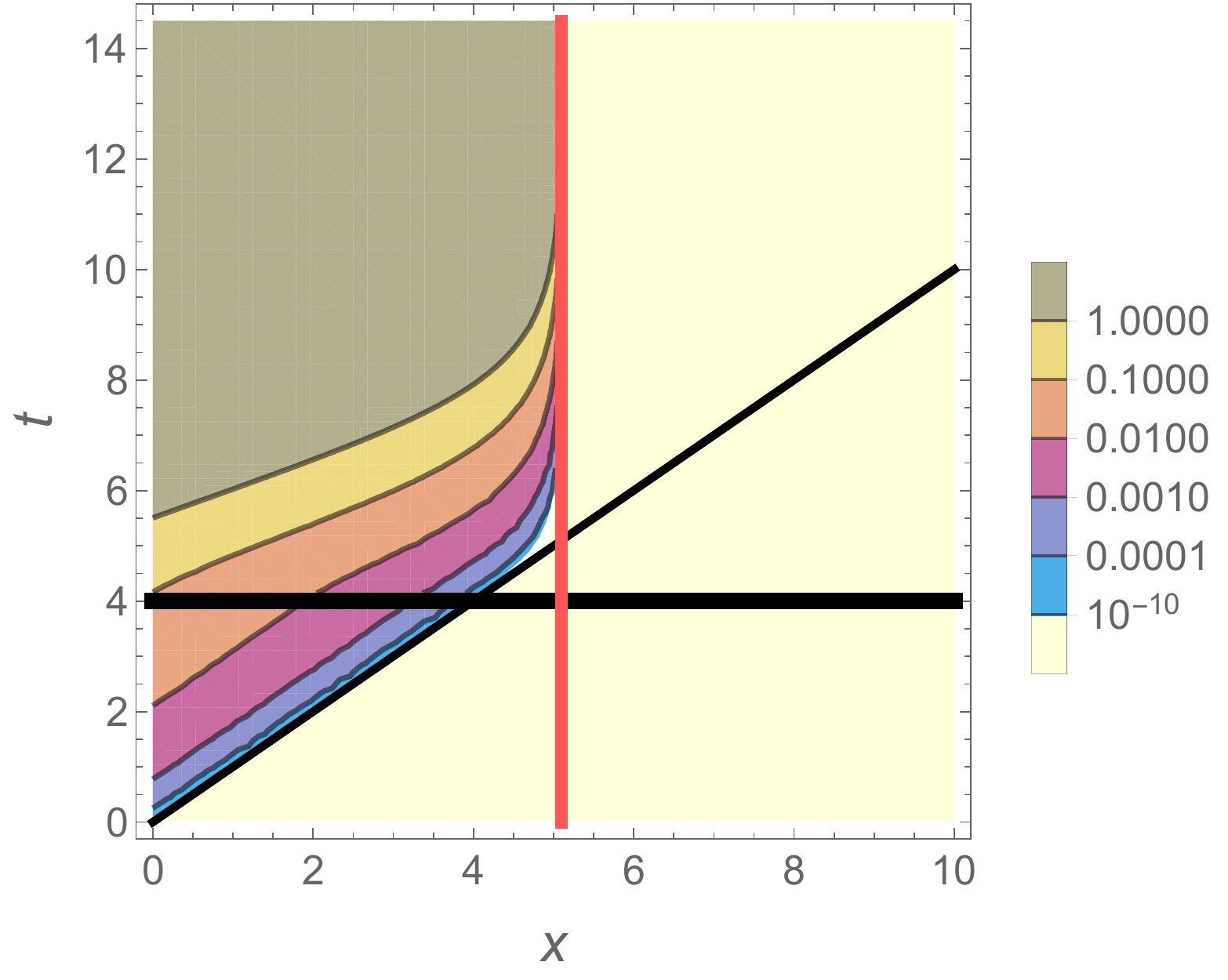}
	\caption{Contour plot of one of the contributions to the eikonal phase shift $\delta(t,x)$ in BTZ-Vaidya, for $G_N= 5*10^{-6},\ m_V=m_W=1,\ \varepsilon=0.1,\ r_-=1,\ r_+=2$. The quench is applied at $t=v_s=4$ and is indicated by the black horizontal line. There is a maximal transverse direction $x_{max}$ indicated by the red vertical line, on which all contour lines accumulate and beyond which this contribution is trivial.}
	\label{fig:accumulation}
\end{figure}

In Figs.~\ref{fig:BTZVaidya}-\ref{fig:AdSVaidya} of Section~\ref{section:results}, it has been noticed that contour lines are bending upwards close to the hatched regions, where we do not have the appropriate formula describing the interaction of the outgoing geodesic with the gravitational shock wave produced by the ingoing geodesic. However, we have a formula for the other contribution to the eikonal phase shift \eqref{eikonal general}, namely the one corresponding to the interaction of the ingoing geodesic with the gravitational shock wave produced by the outgoing geodesic, given by
\begin{equation}
\frac{1}{4}\int h_{\mu\nu}^WT^{\mu\nu}_V.
\end{equation}
Since this interaction always occurs in BTZ$_-$, it is given by half of \eqref{eikonal phase BTZ}. In Fig.~\ref{fig:accumulation}, we plot contour lines associated to this known contribution. Strikingly, there is a transverse distance $|x_{max}|$ on which all contour lines accumulate. Indeed, \eqref{u saddle} implies that the position $\bar{u}$ of the outgoing geodesic is bounded,
\begin{equation}
\bar{u}<\bar{u}_{max}\equiv v_s-\frac{1}{r_-}\ln \frac{r_+-r_-}{r_++r_-},
\end{equation}
such that \eqref{causality uv} implies the existence of a maximal transverse separation $|x_{max}|=\bar{u}_{\max}$ beyond which this contribution is trivial. Although $x_{max}$ lies in the hatched regions of Figs.~\ref{fig:BTZVaidya}-\ref{fig:AdSVaidya}, part of the accumulation behavior is retained in the form of upward bending of contour lines. As a final comment, although both contributions to the eikonal phase are identical outside of the hatched regions as they both correspond to gravitational interactions occurring in BTZ$_-$, we may expect them to largely differ inside the hatched regions.


\begin{thebibliography}{99}
\bibitem{Sekino:2008he}
Y.~Sekino and L.~Susskind,
{\em ``Fast Scramblers,''}
JHEP {\bf 0810} (2008) 065,
\arXiv{0808.2096} [hep-th].

\bibitem{Maldacena:2015waa}
J.~Maldacena, S.~H.~Shenker and D.~Stanford,
{\em ``A bound on chaos,''}
JHEP {\bf 1608} (2016) 106,
\arXiv{1503.01409} [hep-th].	


\bibitem{Shenker:2013pqa}
S.~H.~Shenker and D.~Stanford,
{\em ``Black holes and the butterfly effect,''}
JHEP {\bf 1403} (2014) 067,
\arXiv{1306.0622} [hep-th].  


\bibitem{Shenker:2013yza}
S.~H.~Shenker and D.~Stanford,
{\em ``Multiple Shocks,''}
JHEP {\bf 1412} (2014) 046
\arXiv{1312.3296} [hep-th].

\bibitem{Roberts:2014isa}
D.~A.~Roberts, D.~Stanford and L.~Susskind,
{\em ``Localized shocks,''}
JHEP {\bf 1503} (2015) 051,
\arXiv{1409.8180} [hep-th].

\bibitem{Roberts:2014ifa}
D.~A.~Roberts and D.~Stanford,
{\em ``Two-dimensional conformal field theory and the butterfly effect,''}
Phys.\ Rev.\ Lett.\  {\bf 115} (2015) no.13,  131603,
\arXiv{1412.5123} [hep-th].

\bibitem{Shenker:2014cwa}
S.~H.~Shenker and D.~Stanford,
{\em ``Stringy effects in scrambling,''}
JHEP {\bf 1505} (2015) 132,
\arXiv{1412.6087} [hep-th].

\bibitem{Kitaev}
A.~Kitaev, 
{\em ``Hidden Correlations in the Hawking Radiation and Thermal Noise,"}
talk given at Fundamental Physics Prize Symposium, Nov. 10, 2014.
Stanford SITP seminars, Nov. 11 and Dec. 18, 2014.

\bibitem{Larkin}
A.~Larkin and Y.~N.~Ovchinnikov, 
{\em ``Quasiclassical method in the theory of superconductivity'',} 
J.~Exp.~Theor.~Phys.~\textbf{28}, 1200-1205, 1969. 


\bibitem{Ruelle}
D.~Ruelle,
{\em ``Resonances of Chaotic Dynamical Systems,''}
Phys.\ Rev.\ Lett.\ {\bf 56} (1986) 405.  

	
\bibitem{Lieb:1972wy}
E.~H.~Lieb and D.~W.~Robinson,
{\em ``The finite group velocity of quantum spin systems,''}
Commun.\ Math.\ Phys.\  {\bf 28} (1972) 251.	

\bibitem{Hastings:2005pr}
M.~B.~Hastings and T.~Koma,
{\em ``Spectral gap and exponential decay of correlations,''}
Commun.\ Math.\ Phys.\  {\bf 265} (2006) 781,
\arXiv{math-ph/0507008} [math-ph].

\bibitem{Mezei:2019dfv}
M.~Mezei and G.~Sárosi,
{\em ``Chaos in the butterfly cone,''}
\arXiv{1908.03574} [hep-th].

\bibitem{Halder:2019ric}
I.~Halder,
{\em ``Global Symmetry and Maximal Chaos,''}
\arXiv{1908.05281} [hep-th].

\bibitem{Birmingham:2001pj}
D.~Birmingham, I.~Sachs and S.~N.~Solodukhin,
{\em ``Conformal field theory interpretation of black hole quasinormal modes,''}
Phys.\ Rev.\ Lett.\  {\bf 88} (2002) 151301,
\arXiv{hep-th/0112055} [hep-th].
	
\bibitem{Banks:1998dd} 
  T.~Banks, M.~R.~Douglas, G.~T.~Horowitz and E.~J.~Martinec,
  {\em ``AdS dynamics from conformal field theory,''}
  \arXiv{hep-th/9808016} [hep-th].

  \bibitem{Balasubramanian:1998de} 
  V.~Balasubramanian, P.~Kraus, A.~E.~Lawrence and S.~P.~Trivedi,
  {\em ``Holographic probes of anti-de Sitter space-times,''}
  Phys.\ Rev.\ D {\bf 59}, 104021 (1999),
  \arXiv{hep-th/9808017} [hep-th].
  
  \bibitem{Balasubramanian:1998sn} 
  V.~Balasubramanian, P.~Kraus and A.~E.~Lawrence,
  {\em ``Bulk versus boundary dynamics in anti-de Sitter space-time,''}
  Phys.\ Rev.\ D {\bf 59}, 046003 (1999),
  \arXiv{hep-th/9805171} [hep-th].
  
  
  \bibitem{Balasubramanian:1999ri} 
  V.~Balasubramanian, S.~B.~Giddings and A.~E.~Lawrence,
  {\em ``What do CFTs tell us about Anti-de Sitter space-times?,''}
  JHEP {\bf 9903}, 001 (1999),
  \arXiv{hep-th/9902052} [hep-th].	
	
	
\bibitem{Balasubramanian:1999zv} 
V.~Balasubramanian and S.~F.~Ross,
{\em ``Holographic particle detection,''}
Phys.\ Rev.\ D {\bf 61}, 044007 (2000),
\arXiv{hep-th/9906226} [hep-th].	
	
\bibitem{videomaldacena} J.~Maldacena, 
{\em ``Chaos and black holes,''}
Harvard Physics Morris Loeb Lectures in Physics, (03/25/16), 
\url{https://www.youtube.com/watch?v=7Dd51agJCcU}.

\bibitem{tHooft:1990fkf}
G.~'t Hooft,
{\em ``The black hole interpretation of string theory,''}
Nucl.\ Phys.\ B {\bf 335} (1990) 138.

\bibitem{Kiem:1995iy}
Y.~Kiem, H.~L.~Verlinde and E.~P.~Verlinde,
{\em ``Black hole horizons and complementarity,''}
Phys.\ Rev.\ D {\bf 52} (1995) 7053,
\arXiv{hep-th/9502074} [hep-th].

\bibitem{Kabat:1992tb}
D.~N.~Kabat and M.~Ortiz,
{\em ``Eikonal quantum gravity and Planckian scattering,''}
Nucl.\ Phys.\ B {\bf 388} (1992) 570,
\arXiv{hep-th/9203082} [hep-th].

\bibitem{Balasubramanian:1995sm} 
  V.~Balasubramanian and H.~L.~Verlinde,
  {\em ``Back reaction and complementarity in (1+1) dilaton gravity,''}
  Nucl.\ Phys.\ B {\bf 464}, 213 (1996),
  \arXiv{hep-th/9512148} [hep-th].

\bibitem{Sfetsos:1994xa}
K.~Sfetsos,
{\em ``On gravitational shock waves in curved space-times,''}
Nucl.\ Phys.\ B {\bf 436} (1995) 721,
\arXiv{hep-th/9408169} [hep-th]. 
     
\bibitem{Aparicio:2011zy}
  J.~Aparicio and E.~Lopez,
  {\em ``Evolution of Two-Point Functions from Holography,''}
  JHEP {\bf 1112} (2011) 082,
  \arXiv{1109.3571} [hep-th].
  
\bibitem{Hubeny:2013dea}
  V.~E.~Hubeny and H.~Maxfield,
  {\em ``Holographic probes of collapsing black holes,''}
  JHEP {\bf 1403} (2014) 097,
  \arXiv{1312.6887} [hep-th].  
  
\bibitem{Balasubramanian:2011ur}
V.~Balasubramanian {\it et al.},
{\em ``Holographic Thermalization,''}
Phys.\ Rev.\ D {\bf 84} (2011) 026010,
\arXiv{1103.2683} [hep-th].

\bibitem{Couch:2019zni}
J.~Couch, S.~Eccles, P.~Nguyen, B.~Swingle and S.~Xu,
{\em ``The Speed of Quantum Information Spreading in Chaotic Systems,''}
\arXiv{1908.06993} [cond-mat.stat-mech].

\bibitem{Keranen:2014lna}
V.~Keranen and P.~Kleinert,
{\em ``Non-equilibrium scalar two point functions in AdS/CFT,''}
JHEP {\bf 1504} (2015) 119,
\arXiv{1412.2806} [hep-th].

\bibitem{Keranen:2015mqc}
V.~Keranen and P.~Kleinert,
{\em ``Thermalization of Wightman functions in AdS/CFT and quasinormal modes,''}
Phys.\ Rev.\ D {\bf 94} (2016) no.2,  026010,
\arXiv{1511.08187} [hep-th]. 

\bibitem{David:2015xqa}
J.~R.~David and S.~Khetrapal,
{\em ``Thermalization of Green functions and quasinormal modes,''}
JHEP {\bf 1507} (2015) 041,
\arXiv{1504.04439} [hep-th]. 

\bibitem{Anous:2016kss}
T.~Anous, T.~Hartman, A.~Rovai and J.~Sonner,
{\em ``Black Hole Collapse in the 1/c Expansion,''
	JHEP {\bf 1607} (2016) 123}
\arXiv{1603.04856} [hep-th].

\bibitem{Anous:2017tza}
T.~Anous, T.~Hartman, A.~Rovai and J.~Sonner,
{\em ``From Conformal Blocks to Path Integrals in the Vaidya Geometry,''}
JHEP {\bf 1709} (2017) 009
\arXiv{1706.02668} [hep-th].

\bibitem{Cruz:1994ir}
N.~Cruz, C.~Martinez and L.~Pena,
{\em ``Geodesic structure of the (2+1) black hole,''}
Class.\ Quant.\ Grav.\  {\bf 11} (1994) 2731,
\arXiv{gr-qc/9401025} [gr-qc].

\bibitem{Stodolsky:1978ks}
L.~Stodolsky,
{\em ``Matter and Light Wave Interferometry in Gravitational Fields,''}
Gen.\ Rel.\ Grav.\  {\bf 11} (1979) 391.

\bibitem{Hollowood:2008kq}
  T.~J.~Hollowood and G.~M.~Shore,
  {\em ``The Causal Structure of QED in Curved Spacetime: Analyticity and the Refractive Index,''}
  JHEP {\bf 0812} (2008) 091,
  \arXiv{0806.1019} [hep-th].
  
\bibitem{Visser:1992pz}
  M.~Visser,
  {\em ``van Vleck determinants: Geodesic focusing and defocusing in Lorentzian space-times,''}
  Phys.\ Rev.\ D {\bf 47} (1993) 2395,
  \arXiv{hep-th/9303020} [hep-th].
  
\bibitem{Poisson:2003nc}
E.~Poisson,
{\em ``The Motion of point particles in curved space-time,''}
Living Rev.\ Rel.\  {\bf 7} (2004) 6,
\arXiv{gr-qc/0306052} [gr-qc].

\bibitem{Kent:2014nya}
C.~Kent and E.~Winstanley,
{\em ``Hadamard renormalized scalar field theory on anti–de Sitter spacetime,''}
Phys.\ Rev.\ D {\bf 91} (2015) no.4,  044044,
\arXiv{1408.6738} [gr-qc].
  
\bibitem{Hotta:1992qy}
  M.~Hotta and M.~Tanaka,
  {\em ``Shock wave geometry with nonvanishing cosmological constant,''}
  Class.\ Quant.\ Grav.\  {\bf 10} (1993) 307. 
  
\bibitem{Podolsky:1997ri}
  J.~Podolsky and J.~B.~Griffiths,
  {\em ``Impulsive gravitational waves generated by null particles in de Sitter and anti-de Sitter backgrounds,''}
  Phys.\ Rev.\ D {\bf 56} (1997) 4756.
  
\bibitem{Horowitz:1999gf}
  G.~T.~Horowitz and N.~Itzhaki,
  {\em ``Black holes, shock waves, and causality in the AdS / CFT correspondence,''}
  JHEP {\bf 9902} (1999) 010
  \arXiv{hep-th/9901012} [hep-th].
  
\bibitem{Afkhami-Jeddi:2017rmx}
N.~Afkhami-Jeddi, T.~Hartman, S.~Kundu and A.~Tajdini,
{\em ``Shockwaves from the Operator Product Expansion,''}
\arXiv{1709.03597} [hep-th]. 
  
\bibitem{Cornalba:2006xk}
  L.~Cornalba, M.~S.~Costa, J.~Penedones and R.~Schiappa,
  {\em ``Eikonal Approximation in AdS/CFT: From Shock Waves to Four-Point Functions,''}
  JHEP {\bf 0708} (2007) 019,
  \arXiv{hep-th/0611122} [hep-th].  
  
\bibitem{Cornalba:2007zb}
L.~Cornalba, M.~S.~Costa and J.~Penedones,
{\em ``Eikonal approximation in AdS/CFT: Resumming the gravitational loop expansion,''}
JHEP {\bf 0709} (2007) 037,
\arXiv{0707.0120} [hep-th].  

\bibitem{DHoker:2002nbb}
E.~D'Hoker and D.~Z.~Freedman,
{\em ``Supersymmetric gauge theories and the AdS / CFT correspondence,''}
\arXiv{hep-th/0201253} [hep-th].

\bibitem{Gradshteyn}
I.~S.~Gradshteyn and I.~M.~Ryzhik,
{\em``Tables of Integrals, Series, and Products,''}
Academic Press, 7th edition (2007).

\end{thebibliography}
\end{document}